\documentclass[acmsmall, authorversion=true, nonacm=true]{acmart}

\usepackage[utf8]{inputenc}

\title{Nudge: Stochastically Improving upon FCFS}

\author{Isaac Grosof}
\email{igrosof@cs.cmu.edu}
\affiliation{
    \institution{Carnegie Mellon University}
    \department{Computer Science Department}
    \city{Pittsburgh}
    \state{PA}
    \country{USA}
}
\author{Kunhe Yang}
\email{yangkunhe19@gmail.com}
\affiliation{%
    \institution{Tsinghua University}
    \department{Institute for Interdisciplinary Information Sciences}
    \city{Beijing}
    \country{China}
}
\author{Ziv Scully}
\email{zscully@cs.cmu.edu}
\affiliation{%
    \institution{Carnegie Mellon University}
    \department{Computer Science Department}
    \city{Pittsburgh}
    \state{PA}
    \country{USA}
}
\author{Mor Harchol-Balter}
\email{harchol@cs.cmu.edu}
\affiliation{%
    \institution{Carnegie Mellon University}
    \department{Computer Science Department}
    \city{Pittsburgh}
    \state{PA}
    \country{USA}
}
\date{February 2021}

\setcopyright{rightsretained}
\acmJournal{POMACS}
\acmYear{2021} \acmVolume{5} \acmNumber{2} 
\acmMonth{6} \acmPrice{}\acmDOI{10.1145/3410220.3460102}

\usepackage{natbib}
\usepackage{graphicx}
\usepackage{amsmath}
\usepackage{amsthm}
\usepackage{epsfig}
\usepackage{footmisc}

\makeatletter
\@ifclassloaded{acmart}{}{
    \usepackage[a4paper,margin=1in]{geometry}
    \usepackage{amssymb}
}
\makeatother

\usepackage{amsfonts,dsfont,mathtools,latexsym,nicefrac,enumerate,xcolor,amstext,color}
\usepackage{hyperref}\hypersetup{
	colorlinks=true,
	linkcolor=blue,
	filecolor=magenta,      
	urlcolor=cyan,
	citecolor=magenta,
} 
\usepackage[capitalize]{cleveref}

\newtheorem{theorem}{Theorem}[section]
\newtheorem{lemma}[theorem]{Lemma}

\theoremstyle{acmdefinition}
\newtheorem{definition}{Definition}[section]

\theoremstyle{definition}
\def\P#1{\mathrm{\bf P}\!\left\{#1\right\}}	
\def\E#1{\mathrm{\bf E}\!\left[#1\right]}
\def\condE#1#2{\mathrm{\bf E}_{#1}\!\left[#2\right]}
\def\alg{\mathsf{Alg}}
\def\fcfs{\mathsf{FCFS}}
\def\tir{\mathsf{TIR}}
\def\asymtir{\mathsf{AsymTIR}}
\def\nudge{\mathsf{Nudge}}
\def\ssmall{S_{\textnormal{\textsf{\tiny small}}}}
\def\slarge{S_{\textnormal{\textsf{\tiny large}}}}
\def\psmall{p_{\textnormal{\textsf{\tiny small}}}}
\def\plarge{p_{\textnormal{\textsf{\tiny large}}}}
\def\swap{\mathsf{swap}}
\def\d{\mathop{}\!\mathrm{d}}
\def\indict#1{\mathbf{1}_{#1}}

\DeclareMathOperator{\Exp}{Exp}
\newcommand{\nop}[2]{\mathsf{NoPoisson}(#1, #2)}
\newcommand{\yestag}{\stepcounter{equation}\tag{\theequation}}

\crefname{equation}{}{}
\Crefname{equation}{Equation}{Equations}

\makeatletter
\theoremstyle{acmplain}
\newtheorem*{rep@theorem}{\rep@title}
\newcommand{\newreptheorem}[2]{%
\newenvironment{rep#1}[1]{%
 \def\rep@title{#2 \ref{##1}}%
 \begin{rep@theorem}}%
 {\end{rep@theorem}}}
\makeatother

\newreptheorem{theorem}{Theorem}
\newreptheorem{lemma}{Lemma}



\begin{document}
\begin{abstract}
    The First-Come First-Served (FCFS) scheduling policy
    is the most popular scheduling algorithm used in practice.
    Furthermore, its usage is theoretically validated:
    for light-tailed job size distributions,
    FCFS has weakly optimal asymptotic tail of response time.
    But what if we don't just care about the asymptotic tail?
    What if we also care about the 99th percentile of response time,
    or the fraction of jobs that complete in under one second?
    Is FCFS still best?
    Outside of the asymptotic regime,
    only loose bounds on the tail of FCFS are known,
    and optimality is completely open.
    
    In this paper, we introduce a new policy, Nudge,
    which is the first policy to provably stochastically improve upon FCFS.
    We prove that Nudge simultaneously improves upon FCFS at \emph{every} point along the tail,
    for light-tailed job size distributions.
    As a result, Nudge outperforms FCFS for every moment and every percentile of response time.
    Moreover, Nudge provides a multiplicative improvement over FCFS in the asymptotic tail.
    This resolves a long-standing open problem by showing that, counter to previous conjecture,
    FCFS is not strongly asymptotically optimal.
\end{abstract}
    
	\maketitle
	
    \renewcommand{\shortauthors}{Isaac Grosof et al.}

	\section{Introduction}
	\label{sec:introduction}
    
    \subsection{The Case for FCFS}
    
	While advanced scheduling algorithms are a popular topic in theory papers, it is unequivocal that the most popular scheduling policy used in practice is still First-Come First-Served (FCFS).  There are many reasons for the popularity of FCFS.  From a practical perspective, FCFS is easy to implement.  Additionally, FCFS has a feeling of being fair.  
	
	However, there are also theoretical arguments for why one should use FCFS.  For one thing, FCFS minimizes the {\em maximum} response time across jobs for {\em any} finite arrival sequence of jobs.
	By {\em response time} we mean the time from when a job arrives until it completes service.

	For another thing, in an M/G/1 with a light-tailed job size distribution, FCFS is known to have a {\em weakly optimal} asymptotic tail of response time \citep{stolyar_largest_2001,boxma_tails_2007}.
	Specifically, using $T$ to denote response time, the asymptotic tail under FCFS is of the form:
	\begin{equation}
	\label{eq:FCFS_asymptotic}
	\P{T^\fcfs > t} \sim C_\fcfs e^{-\theta^* t},    
	\end{equation}
	where ``$\sim$'' indicates that the ratio of the two quantities converges to $1$
	in the $t \to \infty$ limit.
	
	The exponent $\theta^*$ in \eqref{eq:FCFS_asymptotic} is known to be optimal, while the optimality of 
	$C_\fcfs$ is an open problem \citep{boxma_tails_2007}.
	The asymptotic tail growth under FCFS has been compared with more sophisticated policies \citep{boxma_tails_2007}.  It has been shown that, for light-tailed job size distributions, the tail of response time under Processor-Sharing (PS), Preemptive Last-Come-First-Served (PLCFS), and Shortest-Remaining-Processing-Time (SRPT) each take the asymptotic form of 
	$$\P{T > t} \sim C' e^{-\theta' t},$$
	where $\theta'$ is the \emph{worst possible}
	exponential decay rate \citep{nair_tail_2010} over all work-conserving scheduling policies.
	Roughly, FCFS's tail exponent $\theta^*$ arises from the tail of the workload distribution,
	while the other policies' tail exponent $\theta'$ arises from the tail of the busy period distribution,
	which is much larger under light-tailed job size distributions.

	\subsection{The Case For Light-Tailed Job Size Distributions}

	In this paper, we choose to focus on the case of light-tailed job size distributions.  Light-tailed job size distributions show up naturally in workloads where all the transactions are of the same type (say shopping); while there is some variability in the time it takes to purchase an item, even high-variability distributions that arise in such settings are often light-tailed. Also, many natural distributions, like the Normal distribution, Exponential distribution, and all Phase-type distributions, are light-tailed.  Finally, while heavy-tailed job size distributions are certainly prevalent in empirical workloads (see for example \cite{Eurosys20,h-t,CrovellaTaqquBestavros98}), in practice, these heavy-tailed workloads are often {\em truncated}, which immediately makes them light-tailed.  Such truncation can happen because there is a limit imposed on how long jobs are allowed to run.  Alternatively, truncation can occur when a heavy-tailed job size distribution is divided into a few size classes as in \cite{JACM02,JPDC99}
	where the smaller size classes end up being truncated distributions.
	
	\subsection{The Case for Non-Asymptotic Tails}
	Within the world of light-tailed job size distributions, FCFS is viewed as the best policy.
	However, while FCFS has a weakly optimal {\em asymptotic} tail,
	it is not best at minimizing $\P{T > t}$ for {\em all} $t$.
	In practice, one cares less about the asymptotic case than about particular $t$ \cite{QUESTA21}.
	For example, one might want to minimize the fraction of response times that exceed $t = 0.5$ seconds, 
	because such response times are noticeable by users.  One might also want to meet several additional 
	Service Level Objectives (SLOs) where one is charged for exceeding particular response time values,
	such as $t = 1$ minute, or $t = 1$ hour. 
	SLOs are very common in the computing literature \citep{MogulWilkes19,chen_sla_2007,QUESTA21},
	in service industries \cite{davis_how_1991,so_price_1998,urban_establishing_2009},
	and in healthcare \citep{blake_analysis_1996,horwitz_us_2010}.
	Unfortunately, different applications have different SLOs.
	This leads us to ask:
	\begin{quote}
	    {\em When considering $\P{T > t}$, is it possible to strictly improve upon FCFS for \underline{all} values of $t$?}
	\end{quote}
	We are motivated by the fact that, for lower values of $t$, Shortest-Remaining-Processing-Time (SRPT) is better than FCFS, although FCFS clearly beats SRPT for higher values of $t$,
	as FCFS is weakly asymptotically optimal while SRPT is asymptotically pessimal \citep{nair_tail_2010,nuyens_preventing_2008}.  SRPT also minimizes mean response time \cite{Schr68},
	which is closely related to lower values of $t$.  This motivates us to consider whether prioritizing small jobs might have some benefit, even in the world of light-tailed job size distributions.
	
	We ask more specifically:
	\begin{quote}
		{\em Can partial prioritization of small jobs lead to a strict improvement over FCFS? Specifically, is there a scheduling policy which strictly improves upon FCFS with respect to $\P{T > t}$, for {\em every} possible $t$ including large $t$?}
	\end{quote}

	\subsection{Our Answer: Nudge}
	
	\begin{figure}[t]
		\includegraphics[height=0.8in]{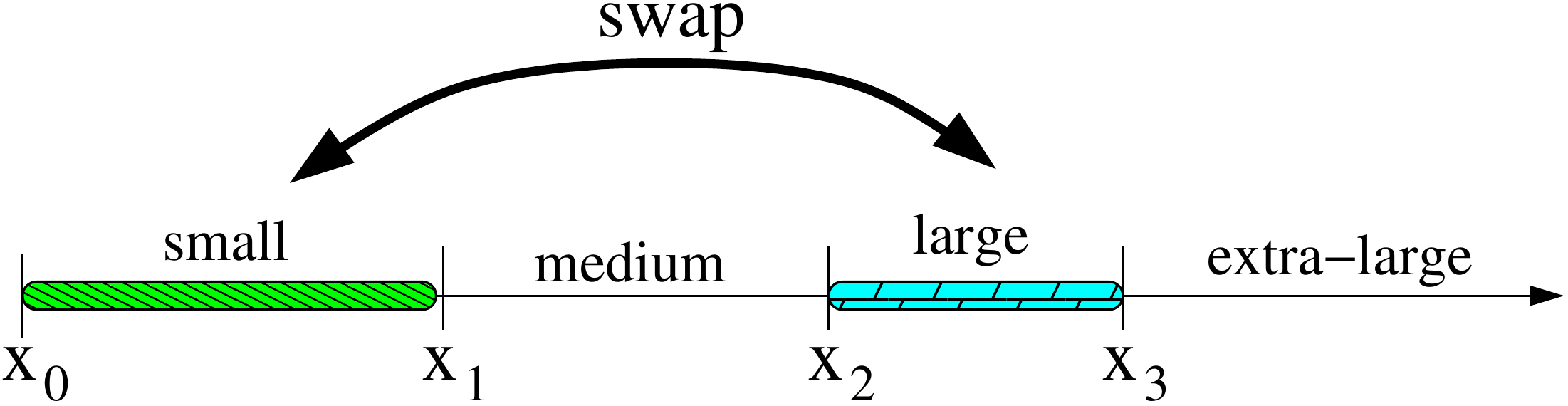}
		\caption{The Nudge algorithm.}
		\label{fig:nudge}
	\end{figure}
	
	\begin{figure}[t]
	    \centering
	    \includegraphics[width=\linewidth]{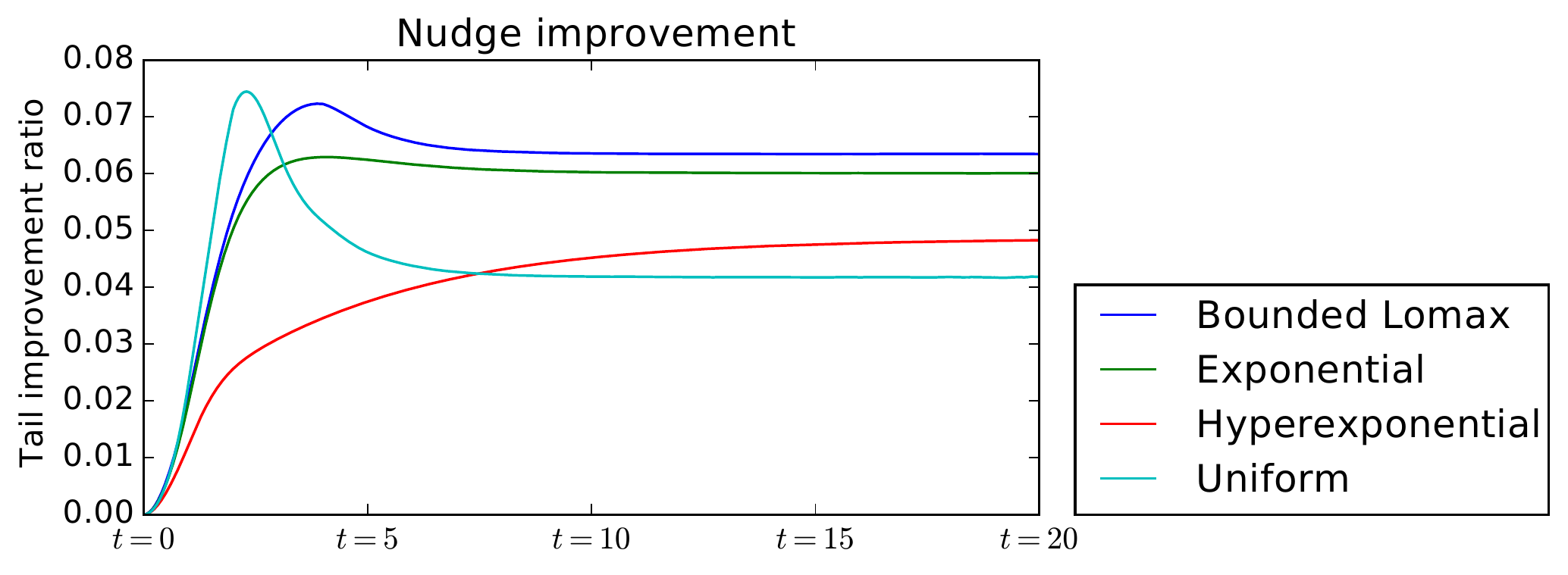}
	    \caption[Text]{Empirical tail improvement of Nudge over FCFS in an M/G/1. The tail improvement ratio (TIR) is defined as $1 - \P{T^\nudge > t}/\P{T^\fcfs > t}$.
	    Specific job size distributions, each with mean $1$: Uniform($0$,$2$); Exponential with mean $1$; Hyperexponential with branches drawn from Exp($2$) and Exp($1/3$), where the first branch has probability $0.8$ and where $C^2 = 3$; BoundedLomax($\lambda=2$, max$=4$, $\alpha = 2$).
	    Distributions in legend ordered by asymptotic improvement.
	    Simulations run with $10$ billion arrivals. Load $\rho = 0.8$. 
	    Nudge parameters:
	    $x_1 = 1, x_2 = 1, x_3 = \infty$.
	    Note that $x_1 = x_2$, so there are no medium-sized jobs.
	    Empirically, Nudge often achieves its best performance with $x_1 = x_2$,
	    though our proofs involve setting $x_1 < x_2$.
	    See \cref{sec:empirical}.
	    }
        
	    \label{fig:improvement_plot}
	\end{figure}

	This paper answers the above question in the affirmative.  We will define a policy, which we call {\em Nudge}, whose response time tail is {\em provably} better than that of FCFS for {\em every} value of $t$,
	assuming a light-tailed job size distribution\footnote{%
        Technically, a Class I job size distribution.
    	See \cref{def:class_I}.}
    (see \cref{thm:existence_of_improvement}).
	We say that Nudge's response time {\em stochastically improves upon} that of FCFS, in the sense of stochastic dominance.
	Moreover, we prove that the asymptotic tail of response time of Nudge is of the form
	\begin{align*}
	    P\{T^\nudge > t\} \sim C_\nudge e^{-\theta^* t},
	\end{align*} with optimal decay rate $\theta^*$
    and a superior leading constant $C_\nudge < C_\fcfs$ (see \cref{cor:existence_of_asymptotic_improvement}).
	Thus, we demonstrate that FCFS is \emph{not} strongly optimal, answering an open problem posed by Boxma and Zwart \citep{boxma_tails_2007}.
	In particular, this is contrary to a conjecture of \citet{wierman_is_2012}
	(see \cref{sub:prior_asymptotic}).
	
	The intuition behind the Nudge algorithm is that we'd like to basically stick to FCFS, which we know is great for handling the extreme tail (high $t$), while at the same time incorporating a little bit of prioritization of small jobs, which we know can be helpful for the mean and lower $t$.   
	We need to be careful, however, not to make too much use of size, because Nudge still needs to beat FCFS for high $t$; hence we want just a little ``nudge'' towards prioritizing small jobs.
	
	We now describe the Nudge algorithm.  Imagine that the job size distribution is divided into size regions, as shown in \cref{fig:nudge}, consisting of small, medium, large, and extra large jobs.  Most of the time, Nudge defaults to FCFS.  However, when a ``small'' job arrives and finds a ``large'' job immediately ahead of it in the queue, we swap the positions of the small and large job in the queue.  The one caveat is that a job which has already swapped is ineligible for further swaps.
	The size cutoffs defining small and large jobs will be defined later in the paper.
	
	The degree of the tail improvement of Nudge over FCFS is non-trivial.
	In \cref{fig:improvement_plot}, we see that for many common light-tailed job size distributions,
	Nudge results in a multiplicative improvement of 4-7\% throughout the tail.
	In \cref{sec:empirical_low_load},
	we show that with low load and a high-variability job size distribution,
	Nudge's improvement can be as much as 10-15\% throughout the tail.
	The magnitude of these improvements highlights the importance of scheduling,
	even in the light-tailed setting.
	
	We additionally present an exact analysis of the performance of Nudge.
	Nudge does not fit into any existing framework for M/G/1 transform analysis,
	including the recently developed SOAP framework \citep{scully_soap_2018} (see \cref{sub:prior_transform}).
	Nonetheless, we derive a tagged-job analysis of Nudge in \cref{thm:transform}, deriving the Laplace-Stieltjes transform of response time of Nudge.

	\subsection{Contributions and Roadmap}
	
	This paper makes the following contributions.
    \begin{itemize}
        \item In \cref{sec:nudge_definition}, we introduce the Nudge policy.
        \item In \cref{sec:main_results,sec:proof_improvement_regime,sec:proof_of_existence_of_improvement} we prove that with appropriately chosen parameters, Nudge stochastically improves upon FCFS for light-tailed%
        \footnote{%
        Technically, any Class I job size distribution.
    	See \cref{def:class_I}.}
    	job size distributions;
        we also give a simple expression for such parameters.
        Moreover, in \cref{sec:proof_of_asymptotic_constant}, we prove that Nudge achieves a multiplicative asymptotic improvement over FCFS.
        \item In \cref{sec:transform}, we derive the Laplace-Stieltjes transform of response time under Nudge.
        \item In \cref{sec:empirical}, we empirically demonstrate the magnitude of Nudge's stochastic improvement over FCFS. We also discuss how to tune Nudge's parameters for best performance.
        \item In \cref{sec:nudge_in_practice},
        we discuss practical considerations for using Nudge.
        \item In \cref{sec:nudge_variants},
        we discuss several notable variants of Nudge.
    \end{itemize}

	\section{Prior Work}
	Most prior work on scheduling to optimize the tail of response time
	focuses on the \emph{asymptotic} case,
	characterizing $\P{T > t}$ in the $t \to \infty$ limit.
	We review these results in \cref{sub:prior_asymptotic}.
	
	Our main result, \cref{thm:existence_of_improvement},
	is a \emph{non-asymptotic} statement,
    characterizing the behavior of $\P{T > t}$ for \emph{all}~$t$,
    not just the $t \to \infty$ limit.
    There is much less prior work on the tail of response time outside of the asymptotic regime.
    We review the few results in this area in \cref{sub:prior_non-asymptotic}.
    
    In addition to characterizing Nudge's tail of response time,
    we also give an exact transform analysis of Nudge's response time.
    Our analysis requires a novel approach
    that significantly differs from traditional analyses,
    as we discuss in \cref{sub:prior_transform}.
    
    Our paper's focus is the M/G/1 queue.
    All of the results cited in this section apply to the M/G/1,
    with some also applying to more general models, such as the GI/GI/1.

	\subsection{Asymptotic Tails: Extensive Theory, but Open Problems Remain}
	\label{sub:prior_asymptotic}
	
    When optimizing the asymptotic tail, the goal is to find a policy $\pi^*$
    such that for all scheduling policies $\pi$,
	\begin{equation*}
	     \limsup_{t \to \infty} \frac{\P{T^{\pi^*} > t}}{\P{T^\pi > t}} \le c
	\end{equation*}
	for some constant~$c \in [1,\infty)$.
	Such a policy $\pi^*$ is called \emph{weakly optimal};
	if $c = 1$, then $\pi^*$ is called \emph{strongly optimal} \citep{boxma_tails_2007}.
	While weak optimality has been well studied,
	proving or disproving strong optimality is much harder.
	
	One major theme of the prior work is that
	optimizing the asymptotic tail looks very different
	depending on the job size distribution.
	\begin{itemize}
	    \item For light-tailed job sizes, FCFS is weakly optimal \citep{boxma_tails_2007}.
	        Specifically, the tail of response time has a form given in \eqref{eq:FCFS_asymptotic}.
	        Moreover, many popular preemptive policies such as PS, SRPT, and Foreground-Background (FB)%
	        \footnote{%
	        FB at all times serves the jobs that have received the least service so far.
	        }
	        are ``weakly pessimal'':
	        they have the \emph{maximum} possible asymptotic tail,
	        up to a constant factor,
	        of any work-conserving scheduling policy \citep{nair_tail_2010,nuyens_preventing_2008}.
	    \item For heavy-tailed job sizes, the reverse is true:
	    PS, SRPT, FB, and similar policies are all weakly optimal
	    \citep{nuyens_preventing_2008,scully_characterizing_2020,boxma_tails_2007},
	    while FCFS is weakly pessimal.
	\end{itemize}
	This state of affairs prompts a question:
	is any policy weakly optimal in both the light-tailed and heavy-tailed cases?
	\Citet{nair_tail_2010} show that a variant of PS achieves this,
	but their variant requires knowledge of the system's load.
	\Citet{wierman_is_2012} show that {\em any} policy that is weakly optimal in both the light- and heavy-tailed cases requires knowing some information about the system parameters, such as the load.
	
	The above results mostly characterize weakly optimal scheduling policies,
	but the problem of strongly optimizing the tail remains open.
	\Citet{boxma_tails_2007} pose the strong optimality of FCFS
	as an open problem.
	\Citet{wierman_is_2012} go further and conjecture that FCFS is in fact
	strongly optimal for light-tailed job size distributions.
	Despite a large of body of work analyzing the tail asymptotics of FCFS
	\citep{abate1994waiting,sakurai2004approximating,abate1997asymptotics,abate1995exponential},
	the problem has remained open.
	We solve the problem by showing that FCFS is \emph{not} strongly optimal.
	Specifically, our \cref{cor:existence_of_asymptotic_improvement} implies
	\begin{align*}
	    \lim_{t \to \infty} \frac{\P{T^\nudge > t}}{\P{T^\fcfs > t}} = \frac{C_\nudge}{C_\fcfs} < 1.
	\end{align*}

	\subsection{Non-asymptotic Tails: Few Optimality Results}
	\label{sub:prior_non-asymptotic}
	
	Characterizing $\P{T > t}$ outside the asymptotic regime
	is a much harder problem than characterizing the asymptotic tail.
	As such, the strongest results in this area are for relatively simple scheduling policies.
	For FCFS under light-tailed job size distributions,
	it is known that $\P{T > t} < e^{-\theta^* t}$ for the same constant $\theta^*$
	as appears in FCFS's asymptotic tail formula \citep{kingman_martingale_1964,kingman_inequalities_1970}.
	As a result, this bound is tight up to a constant factor \citep{kingman_inequalities_1970},
	subject to subtleties discussed in \cref{sec:model_class_I}.
	Beyond FCFS, one of the few known results gives an improved characterization
	of response time under preemptive-priority scheduling policies
	\citep[Section~2]{abate1997asymptotics}.
	
    Very little is known about more complicated scheduling policies.
	While the Laplace-Steiltjes transform of~$T$ is known for
	a wide variety of scheduling policies
	\citep{scully_soap_2018,stanford2014waiting},
	these transforms do not readily yield useful bounds on $\P{T > t}$
	for general job size distributions.
	
	Given that characterizing $\P{T > t}$ is difficult,
	it comes as no surprise that optimizing $\P{T > t}$ is harder still.
	As such, rather than trying to crown a single optimal policy,
	we focus on a \emph{relative} measure.
	Specifically, as we define in \cref{def:stochastic_improvement},
	we say that policy~$\pi_1$ \emph{stochastically improves upon} another policy~$\pi_2$
	if $\P{T^{\pi_1} > t} \leq \P{T^{\pi_2} > t}$ for all~$t$.
	
	There are two stochastic improvement results in the literature,
	but both are much simpler than our Nudge result.
	Both results start with a well-known policy that does not use job sizes
	and show that a variation that does use job sizes stochastically improves response time.
	\begin{itemize}
	    \item \Citet{nuyens_preventing_2008} show that SRPT and similar policies
	    stochastically improve upon FB.
	    \item \Citet{friedman_fairness_2003} and \citet{friedman2003protective} show that
	    one can stochastically improve upon PS
	    by using job sizes.
	    Their policy, Fair Sojourn Protocol (FSP),
	    guarantees in a sample-path sense
	    that no job departs later than it would if the server were using PS.
	\end{itemize}
	
    The results above fit a common theme.
	Both FB and PS often share the server between multiple jobs.
	Sharing the server is fundamentally suboptimal.
	For example, when sharing the server between jobs~1 and~2,
	if we knew that we would finish job~1 first,
	then it would be better to devote the entire server to job~1 at first.
	Doing so improves the response time of job~1 without harming the response time of job~2.
	Roughly speaking, when FB and PS would share the server between jobs,
	SRPT and FSP serve the jobs one at a time,
	using job size information to choose the ordering.
    
    FCFS is more difficult to stochastically improve upon than FB and PS.
    For one thing, FCFS never shares the server,
    removing this easy opportunity for stochastic improvement.
    Moreover, there is a sense in which FCFS is unimprovable:
    on any specific finite arrival sequence,
    FCFS minimizes the sorted vector of response times,
    where we order vectors lexicographically.
    For example, FCFS minimizes the maximum response time.
    As a result, the sample path arguments that work for improving FB and PS
    do not apply to improving FCFS.
    
    In spite of these obstacles, we show in \cref{thm:existence_of_improvement}
    that Nudge stochastically improves upon FCFS.
    Rather than reasoning in terms of sample paths,
    we take a fundamentally stochastic approach from the beginning.
    See our proofs in \cref{sec:proof_improvement_regime}.

    \subsection{Transform of Response Time: Nudge Needs a Novel Approach}
    \label{sub:prior_transform}
    
    In \cref{thm:transform}, we give a closed-form expression for
    the Laplace-Stieltjes transform of Nudge's response time.
    There has been much prior work on analyzing
    the transform of response time of the M/G/1 under various scheduling policies.
    Some analysis techniques cover a wide variety of scenarios
    \citep{scully_soap_2018,fuhrmann_stochastic_1985}.
    However, as we explain below,
    none of these prior techniques can analyze Nudge.
    
    \emph{\textsf{SOAP Policies.}}
	Policies in the SOAP class, introduced by \citet{scully_soap_2018},
	schedule jobs based on an index calculated from each job's size and attained service\footnote{%
    The index can also depend on certain other characteristics of the job,
        e.g. its class if there are multiple classes of jobs.
        Size and attained service are the attributes relevant to Nudge.}, and their response time can be analyzed via the SOAP framework  \citep{scully_soap_2018}.
    These include SRPT \citep{schrage_queue_1966}, FB \citep{schrage_queue_1967},
    some multi-level processor sharing policies \citep{kleinrock_queueing_1976},
    and certain cases of the Gittins policy \citep{osipova_optimal_2009}.
    Unfortunately, Nudge is not a SOAP policy, so we cannot leverage this analysis method.
    This is because whether Nudge will swap a small job~$s$ with a large job~$\ell$
    depends in part on whether any other jobs arrive between $\ell$ and~$s$.
    In contrast,
    a SOAP policy would make such a decision based on properties of $\ell$ and~$s$ alone.
    
    \emph{\textsf{Variations on FCFS.}}
    Nudge serves jobs in FCFS order by default
    and only ever swaps adjacent arrivals.
    One might therefore hope that Nudge could be analyzed as a variation on FCFS.
    There are many papers analyzing a variety of M/G/1 variants under FCFS scheduling.
    These include systems with generalized vacations \citep{fuhrmann_stochastic_1985}
    and exceptional first service \citep{welch_generalized_1964}.
    Unfortunately, to the best of our knowledge,
    no prior analysis of a variation of FCFS applies to Nudge.
    
    \emph{\textsf{Other Analysis Techniques.}}
    There are a number of scheduling policies whose transform analyses
    do not fit into either of the previous categories,
    such as random order of service \citep{kingman_queues_1962}
    and systems with accumulating priority \citep{stanford2014waiting}.
    However, these policies do not resemble Nudge,
    and the techniques used in their analyses do not readily apply to Nudge.

	\section{Model}

	\subsection{Notation}
	
	We consider the M/G/1 queue in which job sizes are known.
	Let $\lambda$ be the arrival rate,
	$S$ be the job size distribution,
	and $s_{\min}$ be the minimum possible job size.
	Specifically, let $s_{\min}$ be the infimum of the support of $S$.
	We denote the load by $\rho=\lambda \E{S}$ and assume $0 < \rho < 1$.
	
	The queueing time, $T_Q$, is the time from when a job arrives until it first receives service. The response time, $T$, is the time from when a job arrives until it completes.
	We write $T_Q^{\alg}$ and $T^{\alg}$ for the queueing time and response time under scheduling algorithm $\alg$, respectively.
	
	For any continuous random variable $V$, we will use $f_V(\cdot)$ to denote the probability density function (p.d.f.) of $V$.   We write the $\widetilde{V}(s)$  for the Laplace-Stieltjes transform of $V$.
	\subsection{Stochastic Improvement}
	
	In this paper, our goal is to prove that the $\nudge$ policy stochastically improves upon the $\fcfs$ policy.
	We now define stochastic improvement, along with the related notion of tail improvement ratio.
	
	\begin{definition}[Stochastic Improvement]
	    \label{def:stochastic_improvement}
		For two scheduling algorithms $\alg_1$ and $\alg_2$, we say that $\alg_1$ (strictly) \emph{stochastically improves} upon $\alg_2$ if, for any response time cutoff $t>s_{\min}$, the probability that response time of $\alg_1$ exceeds $t$ is smaller than the probability that $\alg_2$'s response time exceeds $t$, i.e.,
		\begin{equation*}
	        \forall t > s_{\min}, \quad \P{T^{\alg_1} > t } < \P{T^{\alg_2} > t }.
		\end{equation*}
	\end{definition}
	
	\begin{definition}[Tail improvement ratio]
		For any response time cutoff $t$,  the \emph{tail improvement ratio} of $\alg_1$ versus $\alg_2$ at $t$, denoted $\tir(t)$, is defined as
		\begin{equation*}
			\tir(t)\triangleq 1-\frac{\P{T^{\alg_1}>t}}{\P{T^{\alg_2}>t}}.
		\end{equation*}
		The \emph{asymptotic tail improvement ratio}, denoted $\asymtir$, is defined as
		\begin{equation*}
			\asymtir \triangleq
			\liminf_{t\to\infty} \tir(t)
			=1 - \limsup_{t\to\infty} \frac{\P{T^{\alg_1}>t}}{\P{T^{\alg_2}>t}}.
		\end{equation*}
	\end{definition}
	
	\subsection{Class I ``Light-Tailed'' Distributions}
	\label{sec:model_class_I}
	
	In this paper, we focus on job size distributions for which
	the FCFS policy has an asymptotically exponential waiting time distribution.
	This property of FCFS will be crucial for our analysis.
	Prior work has exactly characterized the job size distributions
	for which FCFS has this property.
	These distributions are known as ``class I'' distributions \citep{abate1994waiting,sakurai2004approximating,abate1997asymptotics}.
    
	\begin{definition}[Class I Distribution]
	    \label{def:class_I}
		For a distribution $S$,
		let $-s^*$ be
		the rightmost singularity of $\widetilde{S}(s)$, with $-s^*=-\infty$ if $\widetilde{S}(s)$ is analytic everywhere.
		$S$ is a \emph{class I distribution} if and only if
		$s^*>0$  and  $\widetilde{S}(-s^*)=\infty$.
	\end{definition}

	Class I distributions can roughly be thought of as ``well-behaved''
    light-tailed distributions.
	In contrast, class II distributions, the other class of light-tailed distributions, are very unusual and ``paradoxical'', and rarely occur as job size distributions. 
	
	For our paper, the key property of class I job size distributions
	is that they cause FCFS to have an asymptotically exponential waiting time distribution for all loads \citep{abate1994waiting,abate1995exponential}. However, as shown by \cite{abate1994waiting,abate1995exponential}, the waiting time also exhibits an exponential tail for light load if the job size is class II. For this reason, while we focus only on class I distributions, we believe that our results also hold for class II under light load. 
	In \cref{sec:fcfs_waiting_time}, we characterize the exponential waiting time in more detail.

  	\subsection{Characterizing the FCFS Waiting Time Distribution}
	\label{sec:fcfs_waiting_time}
	In this paper, we care about the exponential tail of
	the FCFS response time distribution.
	It turns out to be simpler to focus on the FCFS waiting time distribution, which is closely related.
	We will make use of two key concepts regarding the waiting time distribution. The first concept is the asymptotic exponential decay rate, as investigated in~\cite{abate1995exponential,boxma_tails_2007}. We refer to this quantity as $\theta^*$ and formally define it to be the negative of the rightmost singularity of $\widetilde{T_Q^\fcfs}$. Based on the Cramer-Lundberg theory, the waiting time distribution $T_Q^\fcfs$ takes an asymptotic exponential tail:
	\begin{equation}
	    \label{eq:notation_asymptotic_cdf}
	    \P{T_Q^{\fcfs}>t}\sim Ce^{-\theta^*t}.
	\end{equation}
	The quantity $\theta^*$ is the least positive real solution to the equation
	\begin{equation*}
		\widetilde{S}(-\theta^*)=\frac{\lambda + \theta^*}{\lambda}.
	\end{equation*}
	We also define the \emph{normalized p.d.f.} to be
	\begin{align}
	    \label{eq:notation_pdf}
		g(t)\triangleq f_{T_Q^\fcfs}(t)\cdot e^{\theta^* t}.
	\end{align}
	Note that \eqref{eq:notation_asymptotic_cdf} relates to the c.d.f. of waiting time,
    while \eqref{eq:notation_pdf} relates to the p.d.f. of waiting time.
	
	We characterize three important properties of the normalized p.d.f.,
	namely its maximum, minimum, and asymptotic limit.
	Let $g_{\max},\ g_{\min},\ g^*$ denote respectively the maximum, minimum and asymptotically limiting values of $g(\cdot)$ over $[0,\infty)$:
	\begin{equation*}
		g_{\max}\triangleq\sup\limits_{t\in[0,\infty)}g(t);\qquad
		g_{\min}\triangleq\inf\limits_{t\in[0,\infty)}g(t); \qquad
		g^* \triangleq \lim_{t \to \infty} g(t).
	\end{equation*}
	The following lemma, proven in \cref{app:normalized_converges},
	implies these quantities are well defined.
    \begin{lemma}
        \label{lem:normalized_converges}
	    Suppose $S$ is a continuous class I job size distribution.
	    For any load~$\rho$,
	    the normalized p.d.f. $g(t)$
	    is bounded above and below by positive constants,
	    and $\lim_{t \to \infty} g(t)$ exists.
	\end{lemma}
    
    The ratio $g_{\max}/g_{\min}$ will be particularly important in our analysis. Intuitively, we can think of the ratio as measuring the deviation of the queueing time $T_Q^\fcfs$ from a perfect exponential distribution.
    The queueing time distribution is exactly an exponential distribution in an $M/M/1$,
    and diverges from an exponential to greater or lesser degree under any class I job size distribution.
    The degree of divergence will show up in our later results.
    
	\subsection{Scheduling Algorithm: Nudge}
	\label{sec:nudge_definition}
	We now formally define the Nudge algorithm.  
	$\nudge$($x_1, x_2, x_3$) first divides jobs into four regions based on their sizes:
	\begin{itemize}
		\item ``small'': $0=x_0 \le S < x_1$.
		\item ``medium'': $x_1 \le S < x_2$.
		\item ``large'': $x_2 \le S < x_3$.
		\item ``very large'': $x_3 \le S < x_4=\infty$.
	\end{itemize}
	
	Throughout the paper, we concentrate mostly on the ``small'' and the ``large'' jobs. For conciseness, we define $\ssmall,\slarge,\psmall,\plarge$ as follows.
	
	\begin{definition}
	    \label{def:small_large}
	     We define $\ssmall$ and $\slarge$ to be the distribution of small and large jobs, respectively.
	     We also define $\psmall$ and $\plarge$ to be the fraction of small and large jobs, respectively.
	     \begin{align*}
	     \ssmall \sim [S|S<x_1],&\qquad
	     \slarge \sim [S|x_2 \le S < x_3]\\
	     \psmall \triangleq \P{S < x_1},&\qquad
	     \plarge \triangleq \P{x_2 \le S < x_3}.
	     \end{align*}
	\end{definition}
	
	To determine which job to serve, $\nudge$ maintains an ordering over jobs
	which have not yet entered service.
	We call this ordering the ``queue''.
	For each job, we also track whether or not it each has already been ``swapped''.
	
	Whenever a job completes, $\nudge$ serves the job at the front of the queue (if any),
	and serves it to completion.
	By default, newly arriving jobs are placed at the back of the queue,
	resulting in $\fcfs$ scheduling by default.
	However, if {\em three conditions} are satisfied, then a ``swap'' is performed. If
	\begin{enumerate}
	    \item the arriving job is a small job, $j_s$,
	    \item the job at the back of queue is a large job, $j_\ell$, and
	    \item the job $j_\ell$ at the back of queue has never been swapped,
	\end{enumerate}
	then $\nudge$ places the small job $j_s$ just ahead of $j_\ell$,
	in the second-to-last position in the queue.  This is called a {\em swap}, and both $j_\ell$ and $j_s$ are now marked as having been ``swapped.''
	
	Because Nudge never swaps the same job twice, a job is only in the last position in the queue
	and eligible to be swapped immediately after it arrives.
	As a result, Nudge only ever swaps a job with the job that arrives immediately before or after it.

	\section{Main Results}
	\label{sec:main_results}
	
	\subsection{Nudge Improves upon FCFS Non-Asymptotically}
	\label{sec:nudge_improves_non-asymptotic}
	
	Our main goal is to show that Nudge stochastically improves upon FCFS.
	Nudge's performance crucially depends on the choice of parameters $x_1$, $x_2$, and~$x_3$,
	i.e. which jobs are small and which jobs are large.
	We begin by asking:
	given job size distribution~$S$ and load~$\rho$,
	for what choices of parameters $x_1$, $x_2$, and~$x_3$
	does Nudge stochastically improve upon FCFS?
	We answer this in \cref{thm:improvement_regime},
	which gives sufficient conditions on the parameters
	for Nudge to stochastically improve upon FCFS.
	We prove \cref{thm:improvement_regime} in \cref{sec:proof_improvement_regime}.
	
	\begin{theorem}[Stochastic Improvement Regime]
		\label{thm:improvement_regime}
		Suppose $S$ is a continuous class I job size distribution.
		Then $\nudge(x_1, x_2, x_3)$ stochastically improves upon $\fcfs$
		for any $s_{\min} < x_1\le x_2\le x_3$
		satisfying\footnote{%
		    Recall from \cref{def:small_large} that
		    $\ssmall$ and $\slarge$ depend on $x_1$, $x_2$, and~$x_3$.
		    This applies throughout the paper.}
		\begin{align}
			\bullet\ \ &\frac{g_{\max}}{g_{\min}}
			\frac{\lambda+\theta^*}{\lambda}<
			\frac{1-\widetilde{\slarge}(-\theta^*)^{-1}}
			{1-\widetilde{\ssmall}(-\theta^*)^{-1}},
			\label{eq:gmax/gmin_<_function(x_1,x_2)}\\
			\bullet\ \ &x_1+x_3\le 2x_2.
			\label{eq:x1_x2_x3_to_remove_t}
		\end{align}
	\end{theorem}
	
	With \cref{thm:improvement_regime} in hand,
	our goal reduces to the following question:
	given $S$ and~$\rho$, do there exist parameters
	satisfying the sufficient condition from \cref{thm:improvement_regime}?
	We answer this in \cref{thm:existence_of_improvement},
	showing that as long as the minimum job size $s_{\min} = 0$,
	such parameters always exist.
	Our proof of \cref{thm:existence_of_improvement} in \cref{sec:proof_of_existence_of_improvement}
	gives a simple construction of those parameters.
	
	\begin{theorem}[Existence of Stochastic Improvement]
		\label{thm:existence_of_improvement}
		For any continuous class I job size distribution $S$ with $s_{\min} = 0$ and any load $0<\rho<1$, there exist $x_1, x_2, x_3$ satisfying \eqref{eq:gmax/gmin_<_function(x_1,x_2)} and~\eqref{eq:x1_x2_x3_to_remove_t},
		implying that $\nudge(x_1,x_2,x_3)$ stochastically improves upon $\fcfs$.
	\end{theorem}

	\subsection{Nudge Improves upon FCFS Asymptotically}
	\label{sec:nudge_improves_asymptotic}
	
	Having shown that Nudge stochastically improves upon FCFS,
	we ask: is Nudge's improvement non-negligible in the asymptotic limit?
	We answer this in \cref{thm:asymptotic_constant}.
    Specifically, recall that in the $t \to \infty$ limit,
    $\P{T^\fcfs > t} \sim C_\fcfs e^{-\theta^* t}$.
    We show that $\P{T^\nudge > t} \sim C_\nudge e^{-\theta^* t}$
    and that, with appropriately set parameters,
    $C_\nudge < C_\fcfs$.
    This implies that FCFS is \emph{not} strongly optimal for asymptotic tail behavior (see \cref{sub:prior_asymptotic}),
    resolving a long-standing open problem \citep{boxma_tails_2007,wierman_is_2012}.
    We also exactly derive the difference $C_\fcfs - C_\nudge$.
    We prove \cref{thm:asymptotic_constant} in \cref{sec:proof_of_asymptotic_constant},
    making use of \cref{thm:transform}.
    
    \begin{theorem}[Asymptotic Improvement Regime]
	    \label{thm:asymptotic_constant}
		Suppose $S$ is a continuous class I job size distribution.
		For any $s_{\min} < x_1 \leq x_2 \leq x_3$,
		the asymptotic tail improvement ratio
		of $\nudge(x_1, x_2, x_3)$ compared to $\fcfs$ is
		\begin{equation*}
			\asymtir =
			\psmall\plarge \frac{\lambda}{\lambda+\theta^*}\left(
				\widetilde{\slarge}(-\theta^*)
				-\frac{\lambda}{\lambda+\theta^*}
				\widetilde{\ssmall}(-\theta^*)
				-\frac{\theta^*}{\lambda+\theta^*}
				\widetilde{\slarge}(-\theta^*)\widetilde{\ssmall}(-\theta^*)
			\right).
		\end{equation*}
		Furthermore, $\asymtir$ is positive, meaning $C_\nudge < C_\fcfs$, if
		\begin{equation*}
			\frac{\lambda+\theta^*}{\lambda}<
			\frac{1-\widetilde{\slarge}(-\theta^*)^{-1}}
			{1-\widetilde{\ssmall}(-\theta^*)^{-1}}.
		\end{equation*}
	\end{theorem}
	
    Note that the asymptotic improvement regime in \cref{thm:asymptotic_constant}
    is a superset of the non-asymptotic improvement regime in \cref{thm:improvement_regime},
    because $g_{\max}/g_{\min} \ge 1$.
    Thus, whenever \cref{thm:improvement_regime} guarantees a stochastic improvement,
    we also have $C_\fcfs > C_\nudge$.
    Thus, by \cref{thm:existence_of_improvement},
    there exists an asymptotic improvement whenever $s_{\min} = 0$.
	
	\begin{corollary}[Existence of Asymptotic Improvement]
	    \label{cor:existence_of_asymptotic_improvement}
		For any continuous class I job size distribution $S$ with $s_{\min} = 0$ and any load $0<\rho<1$, there exist $x_1, x_2, x_3$ such that $C_\nudge < C_\fcfs$.
	\end{corollary}
	
	While \cref{thm:asymptotic_constant} shows that there is a multiplicative improvement
	in the asymptotic tail, we find empirically that the same multiplicative improvement
	exists throughout nearly the entire tail.
	See \cref{fig:improvement_plot} and \cref{sec:empirical}.

    \subsection{Exact Analysis of Nudge}
    All of the above results compare Nudge's performance to that of FCFS.
    In particular, none of these results give an exact analysis of Nudge's response time.
    We give such an analysis in \cref{thm:transform},
    in which we exactly derive $\widetilde{T^\nudge}(s)$.
	This result is nontrivial, because Nudge does not fall
	into any class of policies with known analyses (see \cref{sub:prior_transform}).
	We prove \cref{thm:transform} in \cref{sec:transform}.
    
    \begin{theorem}[Transform of Response Time]
	\label{thm:transform}
	    The response time of Nudge has Laplace-Stieltjes transform
	    \begin{align*}
	        \SwapAboveDisplaySkip
	        \widetilde{T^\nudge}(s)
	        &= \widetilde{T^\fcfs}(s)
	        + \psmall \plarge \left(
	            \widetilde{\slarge}(s) (1 - \widetilde{\ssmall}(s))
	                \left(\widetilde{T_Q^\fcfs}(\lambda + s) - \widetilde{T_Q^\fcfs}(s)\right)
	         \rule{0pt}{2.2em} \right. \\[-0.8em] &\qquad \left. \rule{0pt}{2.2em}
	            + \widetilde{\ssmall}(s) (1 - \widetilde{\slarge}(s))
	                \left(\frac{\widetilde{T_Q^\fcfs}(s)}{\widetilde{S}(s)} - (1 - \rho)\frac{\lambda/\widetilde{S}(\lambda) - s/\widetilde{S}(s)}{\lambda - s}\right)
	        \right).
	    \end{align*}
	\end{theorem}
	
    \section{Proof of Theorem~\ref{thm:improvement_regime}: Stochastic Improvement Regime}
    
	\label{sec:proof_improvement_regime}
	
	Our goal in this section is to prove \cref{thm:improvement_regime},
	which gives sufficient conditions on the parameters $x_1$, $x_2$, and~$x_3$
	for Nudge to stochastically improve upon FCFS.
	To do so, we employ a tagged job approach.
	In particular, we follow an arbitrary tagged job~$i$ making its way through a pair of coupled systems,
	one employing the FCFS policy and one employing the Nudge policy,
	both with the same arrival process and job sizes.
	
	We focus on one particular response time threshold~$t$,
	and in particular on the events $D_{i, t}$ and $I_{i, t}$,
	where the tagged job~$i$
	has response time greater than~$t$
	in one system and below in the other system.
	In \eqref{eq:relationship_alg1_vs_fcfs}, we write the difference in the response time tails
	of Nudge and FCFS in terms of the events $D_{i, t}$ and $I_{i, t}$.
	In \cref{lem:calculation_of_PI_and_PD},
	we derive formulas for the probabilities of these events.
	
	Using these formulas, in \cref{lem:single_threshold_improvement},
	we derive a sufficient condition
	for Nudge to improve upon FCFS relative to a specific threshold~$t$.
	This sufficient condition is dependent on the threshold~$t$.
	In order to remove this dependence,
	we prove \cref{lem:A_B_c},
	a technical lemma regarding arbitrary random variables.
	
	Finally, in \cref{sec:final_proof_improvement_regime}, we prove \cref{thm:improvement_regime},
	by demonstrating that the conditions given in \cref{thm:improvement_regime}
	ensure that the sufficient condition in \cref{lem:single_threshold_improvement}
	holds relative to every response time threshold~$t$,
	making use of \cref{lem:A_B_c} to do so.

	\subsection{Intermediate Lemmas}
	\label{sub:proofs_of_lemmas}
	
	Consider a tagged job $i$ that arrives into the steady state of the pair of coupled systems.
	We write $T_i^{\nudge}$ and $T_i^{\fcfs}$
	for job~$i$'s response time in the Nudge and FCFS systems, respectively.
	For any $t \ge 0$, define the events
	\begin{equation*}
		I_{i,t} \triangleq \left\{T_i^{\fcfs} \le t < T_i^{\nudge}\right\},
		\qquad
		D_{i,t} \triangleq \left\{T_i^{\nudge} \le t < T_i^{\fcfs}\right\}.
	\end{equation*}
	Intuitively, $D_{i,t}$ is the event in which
	Nudge \emph{decreases} job~$i$'s response time relative to FCFS,
	specifically from above~$t$ to below~$t$.
	Similarly, $I_{i,t}$ is the event in which
	Nudge \emph{increases} job~$i$'s response time relative to FCFS.
    We can write
    \begin{align}
        \label{eq:relationship_alg1_vs_fcfs}
    	\P{T_i^{\nudge} > t}=\P{T_i^{\fcfs} > t}+\P{I_{i,t}}-\P{D_{i,t}}.
    \end{align}
    
    The events $D_{i,t}$ and $I_{i,t}$ are defined using the Nudge and FCFS systems.
    Our next step is to express them in terms of only the FCFS system,
    which we understand well.
    
    We begin by defining the relevant quantities in the FCFS system.
    Let $i^-$ be the arrival immediately before job~$i$, and let $i^+$ be the arrival immediately after,
    and let
    \begin{align*}
        W_i &\triangleq \text{amount of work in the system (either Nudge or FCFS) when job~$i$ arrives,} \\
        A_i &\triangleq \text{interarrival time between jobs~$i$ and~$i^+$,} \\
        S_i &\triangleq \text{size of job~$i$.}
    \end{align*}
    We define analogous quantities for $i^-$ and~$i^+$.
    The work is the same in both systems because both Nudge and FCFS are work-conserving.
    
    Note that Nudge will only ever swap job~$i$ with one of the adjacent arrivals, $i^-$ or~$i^+$
    (see \cref{sec:nudge_definition}).
    Under what condition do we swap job~$i$ with job~$i^+$?
    This happens if and only if the following events occur:
    \begin{enumerate}
        \item[(a)] Job~$i$ is large, which is when $x_2 \le S_i < x_3$.
        \item[(b)] Job~$i^+$ is small, which is when $S_{i^+} < x_1$.
        \item[(c)] Job~$i^+$ arrives before job~$i$ enters service in the Nudge system.
        \item[(d)] Job~$i$ has not swapped with any other job, namely job~$i^-$.
    \end{enumerate}
    Because job~$i$ cannot be both large and small, (a) implies~(d).
    But (d) implies that (c) happens when $A_i \leq W_i$.
    This is because in the absence of swaps,
    job~$i$ would enter service after $W_i$ time.
    Therefore, the event that job~$i$ swaps with job~$i^+$ is
    \begin{equation}
        \label{eq:swap_event}
        \swap_{i,i^+} \triangleq \{(x_2 \leq S_i < x_3) \ \wedge\ (S_{i^+} \leq x_1) \ \wedge\ (A_i \leq W_i)\}.
    \end{equation}
    Crucially, this definition of $\swap_{i,i^+}$ depends only on quantities in the FCFS system.
    We define $\swap_{i^-,i}$ analogously.
	
	\begin{lemma}[Evaluating $\P{I_{i,t}}$ and $\P{D_{i,t}}$]
		\label{lem:calculation_of_PI_and_PD}
		We have
		\begin{align}
			\P{I_{i,t}}&=\P{\swap_{i,i^+} \ \wedge\ (W_i+S_i \le t < W_i+S_i+S_{i^+})},
			\label{eq:P I_i,t}\\
			\P{D_{i,t}}&=\P{\swap_{i^-,i} \ \wedge\ (W_{i^-}-A_{i^-}+S_i \le t < W_{i^-}-A_{i^-}+S_i+S_{i^-})}.
			\label{eq:P D_i,t}
		\end{align}
	\end{lemma}
	\begin{proof}
	    We begin by computing $\P{I_{i,t}}$.
	    The event $I_{i,t}$ occurs only if $T_i^\nudge > T_i^\fcfs$,
	    which in turn occurs only if job~$i$ swaps with the next arrival, namely job~$i^+$.
	    If this swap occurs, then $T_i^\nudge = T_i^\fcfs + S_{i^+}$.
	    We know that $T_i^\fcfs = W_i + S_i$,
	    so \eqref{eq:P I_i,t} follows from
	    \begin{align*}
	        I_{i, t} &= \swap_{i,i^+} \ \wedge\ (T_i^\fcfs \leq t < T_i^\nudge) \\
	        &= \swap_{i,i^+} \ \wedge\ (T_i^\fcfs \leq t < T_i^\fcfs + S_{i^+}) \\
	        &= \swap_{i,i^+} \ \wedge\ (W_i + S_i \leq t < W_i + S_i + S_{i^+}).
	    \end{align*}
	    
	    We now compute $\P{D_{i,t}}$.
	    By similar reasoning to the above,
	    the event $D_{i,t}$ occurs only if job~$i$ swaps with job~$i^-$.
	    If this swap occurs, then $T_i^\nudge = T_i^\fcfs - S_{i^-}$.
	    We again have $T_i^\fcfs = W_i + S_i$, so
	    \begin{align*}
	        D_{i, t} &= \swap_{i^-,i} \ \wedge\ (T_i^\nudge \leq t < T_i^\fcfs) \\
	        &= \swap_{i^-,i} \ \wedge\ (T_i^\fcfs - S_{i^-} \leq t < T_i^\fcfs) \\
	        &= \swap_{i^-,i} \ \wedge\ (W_i + S_i - S_{i^-} \leq t < W_i + S_i).
	    \end{align*}
	    To obtain~\eqref{eq:P D_i,t},
	    observe that conditioned on $\swap_{i^-,i}$,
	    we have $W_i = W_{i-} + S_{i^-} - A_{i^-}$.
	\end{proof}
	
	Now, we give sufficient conditions for Nudge to improve upon FCFS
	relative to a particular threshold~$t$.
	
	\begin{lemma}[Strict Improvement at a Given Threshold]
		\label{lem:single_threshold_improvement}
		Given any $t>s_{\min}$, where $s_{\min}$ is the smallest value of $S$,
		\begin{equation*}
			\P{T^{\nudge} > t } < \P{T^{\fcfs} > t }
		\end{equation*}
		if the following inequality in terms of $t$ holds:
		\begin{align}
			\frac{g_{\max}}{g_{\min}}\frac{\lambda+\theta^*}{\lambda}
			< 
			\frac{
			    \E{e^{\theta^*\min(t,\ssmall+\slarge)}-
				e^{\theta^*\min(t, \ssmall)}}
			}{
				\E{e^{\theta^*\min(t,\ssmall+\slarge)}-
				e^{\theta^*\min(t, \slarge)}}
			}.
			\label{ineq: improve for single t}
		\end{align}
	\end{lemma}
	
	\begin{proof}
	    Because the tagged job~$i$ is a random sample arriving to the steady state of the system,
	    by \cref{eq:relationship_alg1_vs_fcfs},
	    we have $\P{T^{\nudge}> t} < \P{T^{\fcfs}> t}$ if and only if $\P{I_{i,t}} < \P{D_{i,t}}$.
	    Our approach is to use \cref{lem:calculation_of_PI_and_PD}
	    to bound each of $\P{I_{i,t}}$ and $\P{D_{i,t}}$,
	    from which the desired sufficient condition follows.
	    
		We begin by computing $\P{I_{i,t}}$:
		\begin{align*}
			\P{I_{i,t}}
			&= \P{\swap_{i,i^+} \ \wedge\ (W_i+S_i \le t < W_i+S_i+S_{i^+})}
		\tag*{\small [by \cref{lem:calculation_of_PI_and_PD}]}\\
			&= \P{
				\left(A_i\le W_i\right)\wedge
				\left(W_i+S_i\le t< W_i+S_i+S_{i^+}\right)\wedge
				\left(x_2\le S_i<x_3\right)\wedge
				\left(S_{i^+}<x_1\right)}
		\tag*{\small [by \eqref{eq:swap_event}]}\\
			&\le \P{
				\left(W_i+S_i\le t< W_i+S_i+S_{i^+}\right)\wedge
				\left(x_2\le S_i<x_3\right)\wedge
				\left(S_{i^+}<x_1\right)}
		\tag*{\small [discarding $A_i \leq W_i$]}\\
			&= \P{\left((t-S_i-S_{i^+})^+\le W_i\le (t-S_{i})^+\right)\wedge
				\left(x_2\le S_i<x_3\right)\wedge
				\left(S_{i^+}<x_1\right)}
		\\
			&= \plarge\psmall\cdot
			\condE{S_i\sim \slarge,S_{i^+}\sim \ssmall}{\int_{w=(t-S_i-S_{i^+})^+}^{(t-S_{i})^+}
				f_{W_i}(w)\d w
			}
		\tag*{\small$\biggl[\parbox{13em}{change of measure for $S_i$ and $S_i^+$, independence of $S_i$, $S_{i^+}$, and~$W_i$}\biggr]$}\\
			&\le \plarge\psmall\cdot
			\condE{S_i\sim \slarge,S_{i^+}\sim \ssmall}{\int_{w=(t-S_i-S_{i^+})^+}^{(t-S_{i})^+}
				g_{\max}e^{-\theta^* w}\d w
			}
		\tag*{\small$\raisebox{0.15em}{$\biggl[$}\parbox{8.6em}{by \cref{lem:normalized_converges} and the fact that $W_i \sim T_Q^\fcfs$}\raisebox{0.15em}{$\biggr]$}$}\\
			&= \plarge\psmall\cdot
			\frac{e^{-\theta^* t}}{\theta^*}\cdot g_{\max}
			\E{e^{\theta^*\min(t,\ssmall+\slarge)}-
				e^{\theta^*\min(t, \slarge)}}.
				\yestag \label{eq:I_i_t_bound}
		\end{align*}
		Similarly, we compute $\P{D_{i,t}}$:
		\begin{align*}
			&\P{D_{i,t}}
			= \P{\swap_{i^-,i} \ \wedge\ (W_{i^-}-A_{i^-}+S_i \le t < W_{i^-}-A_{i^-}+S_i+S_{i^-})}
		\\
			&=\P{
				(A_{i^-}\le W_{i^-})\wedge
				(W_{i^-}-A_{i^-}+S_i\le t< W_{i^-}-A_{i^-}+S_i+S_{i^-}) 
				\wedge 
				(S_i<x_1)\wedge
				(x_2\le S_{i^-}<x_3)
				}
		\\
			&=\P{\left(A_{i^-}+(t-S_i-S_{i^-})^+\le W_{i^-}\le A_{i^-}+(t-S_{i})^+\right)\ \wedge\
				\left(S_i<x_1\right)\ \wedge\
				\left(x_2\le S_{i^-}<x_3\right)}
		\\
			&=
			\psmall\plarge\cdot
			\condE{S_i\sim \ssmall,S_{i^-}\sim \slarge,A_{i^-}\sim \Exp(\lambda)}
			{\int_{w=A_{i^-}+(t-S_i-S_{i^-})^+}^{A_{i^-}+(t-S_{i})^+}
				f_{W_{i^-}}(w)\d w
			}
		\tag*{\small$\biggl[\parbox{8.6em}{mutual independence of $S_i$, $S_{i^+}$, $A_{i^-}$, and~$W_i$}\biggr]$}\\
			&\ge
			\psmall\plarge\cdot
			\condE{S_i\sim \ssmall,S_{i^-}\sim \slarge,A_{i^-}\sim \Exp(\lambda)}
			{\int_{w=A_{i^-}+(t-S_i-S_{i^-})^+}^{A_{i^-}+(t-S_{i})^+}
				g_{\min}e^{-\theta^*w}\d w
			}
		\tag*{\small$\raisebox{0.15em}{$\biggl[$}\parbox{8.6em}{by \cref{lem:normalized_converges} and the fact that $W_i \sim T_Q^\fcfs$}\raisebox{0.15em}{$\biggr]$}$}\\
			&\ge
			\psmall\plarge\cdot
			\condE{S_i\sim \ssmall,S_{i^-}\sim \slarge}
			{\int_{a=0}^\infty \int_{w=a+(t-S_i-S_{i^-})^+}^{a+(t-S_{i})^+}
				g_{\min}e^{-\theta^*w} \cdot \lambda e^{-\lambda a} \d w \d a
			}
		\tag*{\small [$A_{i^-} \sim \Exp(\lambda)$]}\\
			&=
			\psmall\plarge\cdot
			\frac{e^{-\theta^* t}}{\theta^*}\cdot
			g_{\min}\frac{\lambda}{\lambda+\theta^*}
			\E{e^{\theta^*\min(t,\ssmall+\slarge)}-
				e^{\theta^*\min(t, \ssmall)}}.
				\yestag \label{eq:D_i_t_bound}
		\end{align*}
		Combining the bounds \eqref{eq:I_i_t_bound} and~\eqref{eq:D_i_t_bound},
		we find that $\P{I_{i,t}} < \P{D_{i,t}}$ holds if
		\begin{equation}
		\label{ineq: improve last step}
		    \mkern-22mu
			g_{\max}
			\E{e^{\theta^*\!\min(t,\ssmall+\slarge)}\!-\!
				e^{\theta^*\!\min(t, \slarge)}}<
			g_{\min}\frac{\lambda}{\lambda\!+\!\theta^*}
			\E{e^{\theta^*\!\min(t,\ssmall+\slarge)}\!-\!
				e^{\theta^*\!\min(t, \ssmall)}}.
			\qedhere
		\end{equation}
	\end{proof}

    Having proven \cref{lem:single_threshold_improvement},
    we have a sufficient condition, namely~\eqref{ineq: improve for single t}, for Nudge to improve upon FCFS at a specific value of~$t$.
    But our goal is to improve upon FCFS for \emph{all} values of~$t$.
    We therefore seek a condition which implies that \eqref{ineq: improve for single t} holds for all~$t$.
    
    We start by simplifying \eqref{ineq: improve for single t}.
    Let $A = e^{\theta^* \ssmall}$, $B = e^{\theta^* \slarge}$, and $c=e^{\theta^* t}$.
    Then \eqref{ineq: improve for single t} becomes
    \begin{equation}
        \label{eq:A_B_c_motivation}
        \frac{g_{\max}}{g_{\min}}\frac{\lambda+\theta^*}{\lambda}
        <
		\frac{\E{\min(AB,c)-\min(A,c)}}{\E{\min(AB,c) - \min(B, c)}}.
    \end{equation}
    Here the only appearance of the specific value of~$t$ is via~$c$.
    Our strategy is to lower-bound the right-hand side of \eqref{eq:A_B_c_motivation}
    by a quantity that does not include~$c$.
    The following lemma, which we prove in \cref{app:normalized_converges},
    helps accomplish this under an additional assumption.
    
	\begin{lemma}
		\label{lem:A_B_c}
		Let $A,B$ be two independent real-valued random variables and $c$ be a fixed constant.
		Suppose $1\le A\le c$ and $A<B$. Under these assumptions, if $\P{B>c}>0$ and 
		$c\E{B}\ge \E{A}\E{B|B>c}$,	then
		\begin{equation*}
			\frac{\E{\min(AB, c)- \min(A, c)}}{\E{\min(AB, c)-\min(B, c)}}
			= \frac{\E{\min(AB, c)-A}}{\E{\min(AB, c)-\min(B, c)}}
			\ge\frac{\E{AB-A}}{\E{AB-B}}.
		\end{equation*}
	\end{lemma}

	\subsection{Main Proof}
	\label{sec:final_proof_improvement_regime}
	\begin{reptheorem}{thm:improvement_regime}
	    Suppose $S$ is a continuous class I job size distribution.
		Then $\nudge(x_1, x_2, x_3)$ stochastically improves upon $\fcfs$
		for any $s_{\min} < x_1\le x_2\le x_3$
		satisfying
		\begin{align}
			\bullet\ \ &\frac{g_{\max}}{g_{\min}}
			\frac{\lambda+\theta^*}{\lambda}<
			\frac{1-\widetilde{\slarge}(-\theta^*)^{-1}}
			{1-\widetilde{\ssmall}(-\theta^*)^{-1}},
			\label{eq:repeat_gmax/gmin_<_function(x_1,x_2)}\\
			\bullet\ \ &x_1+x_3\le 2x_2.
			\label{eq:repeat_x1_x2_x3_to_remove_t}
		\end{align}
	\end{reptheorem}
    
    \begin{proof}
		We prove \cref{thm:improvement_regime} by verifying the condition in \cref{lem:single_threshold_improvement}. For every $t>s_{\min}$, we will show that Inequalities~\eqref{eq:repeat_gmax/gmin_<_function(x_1,x_2)}
		and
		\eqref{eq:repeat_x1_x2_x3_to_remove_t} together imply \eqref{ineq: improve for single t}.
		
		\begin{enumerate}[(i)]
			\item When $s_{\min}< t\le x_2$, the
			denominator of the right hand side of \eqref{ineq: improve for single t} becomes
			\begin{equation*}
				\E{e^{\theta^*\min(t,\ssmall+\slarge)}-
					e^{\theta^*\min(t, \slarge)}}=\E{e^t-e^t}=0.
			\end{equation*}
			Thus \eqref{ineq: improve for single t} is not well defined,
			but \eqref{ineq: improve last step} holds trivially, which is sufficient. 
			
			\item When $x_2< t< x_3$, we let $A=e^{\theta^* \ssmall},\ B=e^{\theta^* \slarge}$ and $c=e^{\theta^* t}$. Then clearly $1\le A\le c$ and $A<B$. 
			By \eqref{eq:repeat_x1_x2_x3_to_remove_t}, we know that
			\begin{align*}
				c\E{B}&=e^{\theta^* t}\E{e^{\theta^*\slarge}}\ge e^{\theta^*(2x_2)}
				\ge e^{\theta^*(x_1+x_3)}\\
				&\ge \E{e^{\theta^* (\ssmall+\slarge)}}
				\ge \E{e^{\theta^* \min(\ssmall+\slarge, t)}}
				\ge \E{A}\E{B|B>c}.
			\end{align*}
			We can therefore apply \cref{lem:A_B_c} to obtain
			\begin{align}
				\frac{\E{e^{\theta^*\min(\ssmall+\slarge, t)}-e^{\theta^* \ssmall}}}
				{\E{e^{\theta^*\min(\ssmall+\slarge, t)}-e^{\theta^* \min(\slarge, t)}}}
				\ge\frac{\E{e^{\theta^*(\ssmall+\slarge)}-e^{\theta^* \ssmall}}}
				{\E{e^{\theta^* (\ssmall+\slarge)}-e^{\theta^* \slarge}}}.
				\label{eq:final_1}
			\end{align}
			
			Moreover, condition \eqref{eq:repeat_gmax/gmin_<_function(x_1,x_2)} implies that
			\begin{align}
				\frac{g_{\max}}{g_{\min}}
				\frac{\lambda+\theta^*}{\lambda}
				<
				\frac{1-\widetilde{\slarge}(-\theta^*)^{-1}}
			    {1-\widetilde{\ssmall}(-\theta^*)^{-1}}
				=\frac{\E{e^{\theta^*(\ssmall+\slarge)}-e^{\theta^* \ssmall}}}
				{\E{e^{\theta^* (\ssmall+\slarge)}-e^{\theta^* \slarge}}}
				\label{eq:final_2}
			\end{align}
			Combining \eqref{eq:final_1} with \eqref{eq:final_2}
			establishes \eqref{ineq: improve for single t}.
			
			\item When $t\ge x_3$, we have
			$\min(t, \ssmall)=\ssmall$ and $\min(t, \slarge)=\slarge$. Therefore, condition~\eqref{eq:repeat_gmax/gmin_<_function(x_1,x_2)}
			is equivalent to \eqref{ineq: improve for single t}.
		\end{enumerate}
		
		Therefore, for every $t>s_{\min}$, we have proven that $\P{T^{\nudge}> t}<\P{T^{\fcfs}> t}$.
	\end{proof}
	
	\section{Proof of Theorem~\ref{thm:existence_of_improvement}: Existence of Stochastic Improvement}
	\label{sec:proof_of_existence_of_improvement}
	
	\begin{reptheorem}{thm:existence_of_improvement}
		For any continuous class I job size distribution $S$ with $s_{\min} = 0$ and any load $0<\rho<1$, there exist $x_1, x_2, x_3$ satisfying \eqref{eq:repeat_gmax/gmin_<_function(x_1,x_2)} and~\eqref{eq:repeat_x1_x2_x3_to_remove_t},
		implying that $\nudge(x_1,x_2,x_3)$ stochastically improves upon $\fcfs$.
	\end{reptheorem}
		\begin{proof}
		We start by constructing $x_1,x_2,x_3$ that satisfy both
		Inequality \eqref{eq:repeat_gmax/gmin_<_function(x_1,x_2)}
		and \eqref{eq:repeat_x1_x2_x3_to_remove_t}. 
		For notational convenience, let $M=\frac{g_{\max}}{g_{\min}}
		\frac{\lambda+\theta^*}{\lambda}$.
		First fix an arbitrary $x_3>0$ and let $x_2=\frac{3}{4} x_3$, then compute $\widetilde{\slarge}(-\theta^*)$ and choose a small enough $x_1$ such that 
		\begin{align}
			x_1<\min\left\{
			-\frac{1}{\theta^*}\ln\!\left(
			1-
			\frac{1-\widetilde{\slarge}(-\theta^*)^{-1}}{M}
			\right),\frac{1}{2}x_3
			\right\}.
			\label{eq:how_to_choose_x1}
		\end{align}
		Clearly, such $x_1 > s_{\min} = 0$ in \eqref{eq:how_to_choose_x1} exists because $M>1$, so we have
		\begin{align}
			\bullet\ \ &x_1+x_3< \frac{3}{2}x_3=2x_2,
			\label{ineq: satisfy condition 1}\\
			\bullet\ \ &
			\frac{1-\widetilde{\slarge}(-\theta^*)^{-1}}
			{1-\widetilde{\ssmall}(-\theta^*)^{-1}}=\frac{1-\widetilde{\slarge}(-\theta^*)^{-1}}
			{1-\E{e^{\theta^*\ssmall}}^{-1}}\ge
			\frac{1-\widetilde{\slarge}(-\theta^*)^{-1}}{1-e^{-\theta^* x_1}}>
			M=\frac{g_{\max}}{g_{\min}}
			\frac{\lambda+\theta^*}{\lambda}.
			\label{ineq: satisfy condition 2}
		\end{align}
		By \cref{thm:improvement_regime}, \eqref{ineq: satisfy condition 1} and \eqref{ineq: satisfy condition 2} together imply that $\P{T^{\nudge} \ge t } < \P{T^{\fcfs} \ge t }$ for every $t > s_{\min}=0$. Therefore, 
		\begin{equation*}
			\forall t > 0,\quad\P{T^{\nudge} > t } < \P{T^{\fcfs} > t }.
			\qedhere
		\end{equation*}
	\end{proof}

	\section{Proof of Theorem~\ref{thm:transform}: Transform of Response Time}
	\label{sec:transform}
	
	In this section we compute an exact formula for $\widetilde{T^\nudge}(s)$.
	The formula holds for arbitrary job size distributions, not just those of class~I.
	
    At a high level, our analysis works by considering two systems experiencing identical arrivals: one using Nudge, and one using FCFS.
    We consider a tagged job arriving to this pair of systems in equilibrium and determine how its Nudge queueing time relates to its FCFS queueing time.
    \begin{itemize}
        \item \emph{Small jobs:} Nudge queueing time is FCFS queueing time, possibly minus a large job's size.
        \item \emph{Large jobs:} Nudge queueing time is FCFS queueing time, possibly plus a small job's size.
        \item \emph{Other jobs:} Nudge queueing time is identical to FCFS queueing time.
    \end{itemize}
    We will determine $\widetilde{T_{Q,\mathsf{small}}^\nudge}(s)$ and $\widetilde{T_{Q,\mathsf{large}}^\nudge}(s)$,
    from which $\widetilde{T^\nudge}(s)$ easily follows.
    
    
    \subsection{Probabilistic Interpretation of the Laplace-Stieltjes Transform}
    
    Before jumping into the Nudge queueing time analysis,
    we recall a probabilistic interpretation of the Laplace-Stieltjes transform.
    
    Let $V$ be a nonnegative random variable.
    Consider a time interval of length~$V$
    and a Poisson process of rate~$s$ that is independent of~$V$.
    We call the increments of the Poisson process ``interruptions''.
    Let $\nop{V}{s}$ be the event that there are no interruptions
    during the time interval.
    Then \citep[Exercise~25.7]{harchol-balter_performance_2013}
    \begin{equation}
        \label{eq:lst_nop}
        \widetilde{V}(s) = \P{\nop{V}{s}}.
    \end{equation}
    
    The interpretation in \eqref{eq:lst_nop} necessarily requires $s \geq 0$.
    Fortunately, formulas we derive using \eqref{eq:lst_nop}
    are still valid for $s < 0$ because Laplace transforms are uniquely defined
    by their value on any bounded interval on the real line
    \citep{chareka_finite-interval_2007}.

    \subsection{Transform for Large Jobs}
    \label{sec:transform_large}
    
    \begin{lemma}
        \label{lem:lst_large}
        The queueing time of large jobs under Nudge has Laplace-Stieltjes transform
        \begin{equation*}
            \widetilde{T_{Q,\mathsf{large}}^\nudge}(s)
            = \bigl(1 - \psmall(1 - \widetilde{\ssmall}(s))\bigr) \widetilde{T_Q^\fcfs}(s)
            + \psmall(1 - \widetilde{\ssmall}(s)) \widetilde{T_Q^\fcfs}(\lambda + s).
        \end{equation*}
    \end{lemma}

    \begin{proof}
    Consider a large tagged job arriving to the pair of systems,
    one using Nudge and the other using FCFS,
    in equilibrium.
    We can think of the job's Nudge queueing time as
    the time it takes to do the following two steps:
    \begin{enumerate}
        \item[(a)]
        We first wait for its FCFS queueing time, namely~$T_Q^\fcfs$.
        \item[(b)]
        If during that $T_Q^\fcfs$ time there has been at least one arrival,
        and if the first such arrival is a small job,
        we then wait for that small job's service, which takes $\ssmall$ time.
        Note that the small job's size is independent of the FCFS queueing time.
    \end{enumerate}
    
    We will use \eqref{eq:lst_nop} to compute $\widetilde{T_{Q,\mathsf{large}}^\nudge}(s)$.
    To that end, we associate each of the Nudge and FCFS systems with
    a Poisson ``interruption'' process of rate~$s$.
    The interruption processes are independent of
    the arrival times and job sizes of each system,
    but they are coupled to each other in the following way.
    At any moment in time when the systems are busy,
    some job~$j$ has been in service for some amount of time~$t$.
    We couple the interruption processes such that
    interruptions occur at the same $(j, t)$ pairs in both systems.
    
    By \eqref{eq:lst_nop}, $\widetilde{T_{Q,\mathsf{large}}^\nudge}(s)$ is the probability that
    no interruptions occur during steps~(a) and~(b).
    We compute this probability by conditioning on the following event:
    \begin{equation*}
        E = \left\{\text{\parbox{18em}{the next arrival after the tagged job is small, and an interruption occurs during its service}}\right\}
	\end{equation*}
	Note that $E$ does not consider whether the next arrival occurs before the tagged job exits the queue.
	Therefore, it is independent of the length $T_Q^\fcfs$ of step~(a).
	
	If $E$ does not occur,
	then there are no interruptions during step~(b).
	Therefore, there are no interruptions if and only if
	there are no interruptions during step~(a).
	By~\eqref{eq:lst_nop},
	this happens with probability $\widetilde{T_Q^\fcfs}(s)$.
	
	If $E$ does occur,
	then an interruption will occur during step~(b)
	if and only if a new job arrives during step~(a).
	That is, by conditioning on~$E$,
	we have predetermined that the next arrival will be small and,
	if it swaps with the tagged job, will cause an interruption.
	Therefore, to avoid interruptions,
	we need to avoid interruptions \emph{and arrivals} during step~(a).
	Merging the arrival and interruption processes yields
	a Poisson process of rate $\lambda + s$,
	so avoiding interruptions corresponds to the event $\nop{T_Q^\fcfs}{\lambda + s}$.
	By~\eqref{eq:lst_nop},
	this happens with probability $\widetilde{T_Q^\fcfs}(\lambda + s)$.
	
    Conditioning on whether $E$ occurs
    and using \eqref{eq:lst_nop} to compute
    $\P{E} = \psmall(1 - \widetilde{\ssmall}(s))$,
    we obtain the desired expression.
    \end{proof}

	\subsection{Transform for Small Jobs}
	\label{sec:transform_small}
	
    \begin{lemma}
        \label{lem:lst_small}
        The queueing time of small jobs under Nudge has Laplace-Stieltjes transform
        \begin{equation*}
            \widetilde{T_{Q, \mathsf{small}}^\nudge}(s)
            = \widetilde{T_Q^\fcfs}(s) \left(1 + \frac{\plarge(1 - \widetilde{\slarge}(s))}{\widetilde{S}(s)}\right)
            - \plarge(1 - \widetilde{\slarge}(s))(1 - \rho) \cdot \frac{\lambda/\widetilde{S}(\lambda) - s/\widetilde{S}(s)}{\lambda - s}.
        \end{equation*}
    \end{lemma}
    
    The analysis of small jobs is more involved than the analysis of large jobs.
    We therefore state and prove several more intermediate results
    before proving \cref{lem:lst_small}.
	
	Consider a small tagged job arriving to the pair of systems,
	one using Nudge and the other using FCFS,
	in equilibrium.
    The main question we need to answer is
    whether the tagged job will swap with a large job
    in the Nudge system.
    Our main insight is that we can tell whether the swap will occur
    by examining just the FCFS system.
    Because we understand FCFS well,
    this makes it relatively simple to tell whether a swap will occur.
    
    \begin{lemma}
        \label{lem:swap_fcfs}
        The small tagged job swaps with a large job in the Nudge system
        if and only if, when it arrives,
        the FCFS system has a nonempty queue whose last job is large.
    \end{lemma}
    
    The proof of \cref{lem:swap_fcfs} follows very similar reasoning
    to our analysis at the start of \cref{sub:proofs_of_lemmas}.
    For completeness, we provide a proof in \cref{app:transform_proofs}.
	
	Thanks to \cref{lem:swap_fcfs},
	we can determine the queueing time of the small tagged job
	by looking at the state of the FCFS system when it arrives.
	We describe the equilibrium state of the FCFS system with the following quantities:
	\begin{itemize}
	    \item $W$: the amount of work in the system.
        \item $N_Q$: the number of jobs in the queue.
        \item $W_{\mathsf{most}}$: the amount of work in the system,
        excluding the last job in the queue if $N_Q \geq 1$.
        \item $S_{\mathsf{last}}$: the size of the last job in the queue,
        or $0$ if $N_Q = 0$.
	\end{itemize}
    Note that these quantities are not independent.
    In particular, $W = W_{\mathsf{most}} + S_{\mathsf{last}}$.
    However, $W_{\mathsf{most}}$ and $S_{\mathsf{last}}$
    are \emph{conditionally} independent given $N_Q \geq 1$.
    
    Armed with \cref{lem:swap_fcfs} and the system state notation,
    we are ready to compute $T_{Q, \mathsf{small}}^\nudge(s)$,
    thus proving \cref{lem:lst_small}.
    Our computation will make use of an additional lemma
    which we state after the proof and prove in \cref{app:transform_proofs}.
    
    \begin{proof}[Proof of \cref{lem:lst_small}]
    Consider the small tagged job arriving to the pair of systems in equilibrium.
    By \cref{lem:swap_fcfs}, its Nudge queueing time is
    \begin{equation*}
        T_{Q, \mathsf{small}}^\nudge
        = \begin{cases}
            W_{\mathsf{most}}
            & \text{if } N_Q = 0
            \\
            W_{\mathsf{most}} + S_{\mathsf{last}} \mathbf{1}(\lnot(x_2 \leq S_{\mathsf{last}} < x_3))
            & \text{if } N_Q \geq 1.
        \end{cases}
    \end{equation*}
    
    Applying \eqref{eq:lst_nop}
    and the conditional independence of $W_{\mathsf{most}}$ and $S_{\mathsf{last}}$ yields
    \begin{align*}
        \widetilde{T_{Q, \mathsf{small}}^\nudge}(s)
        &= \P{\nop{W_{\mathsf{most}}}{s} \ \land\ N_Q = 0} \\
        &\quad + \P{\nop{W_{\mathsf{most}}}{s} \ \land\ N_Q \geq 1 \ \land\ (\nop{S_{\mathsf{last}}}{s} \ \lor\ x_2 \leq S_{\mathsf{last}} < x_3)} \\
        &= \P{\nop{W_{\mathsf{most}}}{s} \ \land\ N_Q = 0} \\
        \yestag \label{eq:lst_small_progress}
        &\quad + \P{\nop{W_{\mathsf{most}}}{s} \ \land\ N_Q \geq 1}
            \cdot \bigl(\widetilde{S}(s) + \plarge(1 - \widetilde{\slarge}(s))\bigr).
    \end{align*}
    
    It remains only to compute the two probabilities in \eqref{eq:lst_small_progress}.
    Let
    \begin{equation}
        \label{eq:lst_small_q_0}
        q \triangleq \P{\nop{W}{s} \ \land\ N_Q = 0} = \P{\nop{W_{\mathsf{most}}}{s} \ \land\ N_Q = 0},
    \end{equation}
    making $q$ the first probability in \eqref{eq:lst_small_progress}.
    We now compute the second probability in \eqref{eq:lst_small_progress} in terms of~$q$.
    First, note that $T_Q^\fcfs$ and $W$ are identically distributed.
    Recalling the conditional independence of $W_{\mathsf{most}}$ and $S_{\mathsf{last}}$,
    we have, using \eqref{eq:lst_nop} throughout,
    \begin{align*}
        \MoveEqLeft
        \widetilde{T_Q^\fcfs}(s) - q
        = \widetilde{W}(s) - q
        = \P{\nop{W}{s} \ \land\ N_Q \geq 1} \\
        &= \P{\nop{W_{\mathsf{most}}}{s} \ \land\ N_Q \geq 1 \ \land\ \nop{S_{\mathsf{last}}}{s}} \\
        \yestag \label{eq:lst_small_q_1}
        &= \P{\nop{W_{\mathsf{most}}}{s} \ \land\ N_Q \geq 1} \cdot \widetilde{S}(s).
    \end{align*}
    Plugging \eqref{eq:lst_small_q_0} and \eqref{eq:lst_small_q_1}
    into \eqref{eq:lst_small_progress} yields
    \begin{equation*}
        \widetilde{T_{Q, \mathsf{small}}^\nudge}(s)
        = q + \bigl(\widetilde{T_Q^\fcfs}(s) - q\bigr)\left(1 + \frac{\plarge(1 - \widetilde{\slarge}(s))}{\widetilde{S}(s)}\right).
    \end{equation*}
    \Cref{lem:lst_small_q} below, proven in \cref{app:transform_proofs},
    computes the value of~$q$,
    yielding the desired result.
    \end{proof}
    
    \begin{lemma}
        \label{lem:lst_small_q}
        Let $q \triangleq \P{\nop{W}{s} \ \land\ N_Q = 0}$.
        We have
        \begin{equation*}
            q
            = \widetilde{T_Q^\fcfs}(\lambda) \cdot \frac{\lambda \widetilde{S}(s) - s \widetilde{S}(\lambda)}{\lambda - s}
            = \frac{1 - \rho}{\widetilde{S}(\lambda)} \cdot \frac{\lambda \widetilde{S}(s) - s \widetilde{S}(\lambda)}{\lambda - s}.
        \end{equation*}
    \end{lemma}

	\subsection{Overall Response Time Transform}
	\begin{reptheorem}{thm:transform}
	    The response time of Nudge has Laplace-Stieltjes transform
	    \begin{align*}
	        \SwapAboveDisplaySkip
	        \widetilde{T^\nudge}(s)
	        &= \widetilde{T^\fcfs}(s)
	        + \psmall \plarge \left(
	            \widetilde{\slarge}(s) (1 - \widetilde{\ssmall}(s))
	                \left(\widetilde{T_Q^\fcfs}(\lambda + s) - \widetilde{T_Q^\fcfs}(s)\right)
	         \rule{0pt}{2.5em} \right. \\[-0.75em] &\qquad \left. \rule{0pt}{2.5em}
	            + \widetilde{\ssmall}(s) (1 - \widetilde{\slarge}(s))
	                \left(\frac{\widetilde{T_Q^\fcfs}(s)}{\widetilde{S}(s)} - (1 - \rho)\frac{\lambda/\widetilde{S}(\lambda) - s/\widetilde{S}(s)}{\lambda - s}\right)
	        \right).
	    \end{align*}
	\end{reptheorem}
	
	\begin{proof}
	    The expression follows by plugging the results of \cref{lem:lst_small,lem:lst_large} into
	    \begin{align*}
	        \widetilde{T^\nudge}(s)
	        &= \psmall \cdot \widetilde{T_{Q,\mathsf{small}}^\nudge}(s) \cdot \widetilde{\ssmall}(s)
	        + \plarge \cdot \widetilde{T_{Q,\mathsf{large}}^\nudge}(s) \cdot \widetilde{\slarge}(s)\\
	        &\quad + (1 - \psmall - \plarge) \cdot \widetilde{T_Q^\fcfs}(s) \cdot \frac{\widetilde{S}(s) - \psmall \widetilde{\ssmall}(s) - \plarge \widetilde{\slarge}(s)}{1 - \psmall - \plarge}
	    \end{align*}
	    and simplifying the resulting expression.
	    One key step is recognizing
	    $\widetilde{T^\fcfs}(s) =  \widetilde{T_Q^\fcfs}(s) \cdot \widetilde{S}(s)$.
	\end{proof}
	
    \section{Proof of Theorem~\ref{thm:asymptotic_constant}: Asymptotic Improvement}
	\label{sec:proof_of_asymptotic_constant}
	\begin{reptheorem}{thm:asymptotic_constant}
		Suppose $S$ is a continuous class I job size distribution.
		For any $s_{\min} < x_1 \leq x_2 \leq x_3$,
		the asymptotic tail improvement ratio
		of $\nudge(x_1, x_2, x_3)$ compared to $\fcfs$ is
		\begin{equation*}
			\asymtir =
			\psmall\plarge \frac{\lambda}{\lambda+\theta^*}\left(
				\widetilde{\slarge}(-\theta^*)
				-\frac{\lambda}{\lambda+\theta^*}
				\widetilde{\ssmall}(-\theta^*)
				-\frac{\theta^*}{\lambda+\theta^*}
				\widetilde{\slarge}(-\theta^*)\widetilde{\ssmall}(-\theta^*)
			\right).
		\end{equation*}
		Furthermore, $\asymtir$ is positive, meaning $C_\nudge < C_\fcfs$, if
		\begin{equation*}
			\frac{\lambda+\theta^*}{\lambda}<
			\frac{1-\widetilde{\slarge}(-\theta^*)^{-1}}
			{1-\widetilde{\ssmall}(-\theta^*)^{-1}}.
		\end{equation*}
	\end{reptheorem}
	
	Below we give a high-level overview of the proof.
	We provide full computations in \cref{app:transform_proofs}.
	
	\begin{proof}[Proof sketch]
	Using the Final Value Theorem, one can show that for $\alg \in \{\nudge, \fcfs\}$,
	\begin{equation*}
	    C_\alg = \frac{1}{\theta^*}\lim_{s \to 0} s\widetilde{T^\alg}(s - \theta^*).
	\end{equation*}
	Combining this with \cref{thm:transform},
	which relates $\widetilde{T^\nudge}(s)$ to $\widetilde{T^\fcfs}(s)$,
	will relate $C_\nudge$ to $C_\fcfs$.
	
	To obtain $\asymtir = 1 - C_\nudge/C_\fcfs$,
	we compute $\lim_{s \to 0} (s\widetilde{T^\fcfs}(s - \theta^*) - s\widetilde{T^\nudge}(s - \theta^*))$
	via \cref{thm:transform}.
	Each non-vanishing term
	has an $sT^\fcfs_Q(s - \theta^*)$ factor.
	Because $\widetilde{T^\fcfs_Q}(s) = \widetilde{T^\fcfs}(s)/\widetilde{S}(s)$,
	we can express $C_\fcfs - C_\nudge$ as a constant times $C_\fcfs$;
	this constant is $\asymtir$.
	\end{proof}

	\section{Empirical Lessons}
	\label{sec:empirical}
	
	This paper proves that Nudge stochastically improves upon FCFS
	under the correct choice of parameters,
	and achieves multiplicative improvement in the asymptotic tail.
	However, there are a few practical questions remaining.
	These questions center around finding Nudge parameters in practice.
	In this section, we demonstrate several practical lessons on
	choosing Nudge parameters.
	\begin{enumerate}
	    \item (\cref{sec:how_set_nudge_parameters}) We find that Nudge typically achieves its greatest improvement over FCFS when the Nudge parameters specify that all jobs are either large or small (i.e. $x_1 = x_2$, $x_3 = \infty$).
	    \item (\cref{sec:empirical_low_load})
	    We find that when load is low, Nudge can dramatically improve upon FCFS (10-20\%) in the common case where job size variability is relatively high (i.e. $C^2 > 1$).
	    When job size variability is lower and load is low,
	    improvement is smaller.
	    \item (\cref{sec:stochastic_asymptotic_connection})
	    We find that
	    the space of parameters that lead Nudge to {\em asymptotically} improve upon FCFS
	    typically also cause Nudge to {\em stochastically} improve upon FCFS.
	    This is serendipitous, because \cref{thm:asymptotic_constant}
	    provides a simple, exact method to check whether given Nudge parameters
	    will achieve asymptotic improvement over FCFS.
	\end{enumerate}
	
	\subsection{All Jobs Should Be Either Large or Small}
	\label{sec:how_set_nudge_parameters}
	
	When evaluating Nudge on common job size distributions,
	we have found that the greatest improvement over FCFS is achieved
	by setting the Nudge parameters such that
	all jobs are either large or small (i.e. $x_1 = x_2$, $x_3 = \infty$),
	with no medium or very large jobs.
	This is a pattern we have seen with great consistency across a variety of job size distributions.

	\begin{figure}[t]
	    \centering
	    \includegraphics[width=\linewidth]{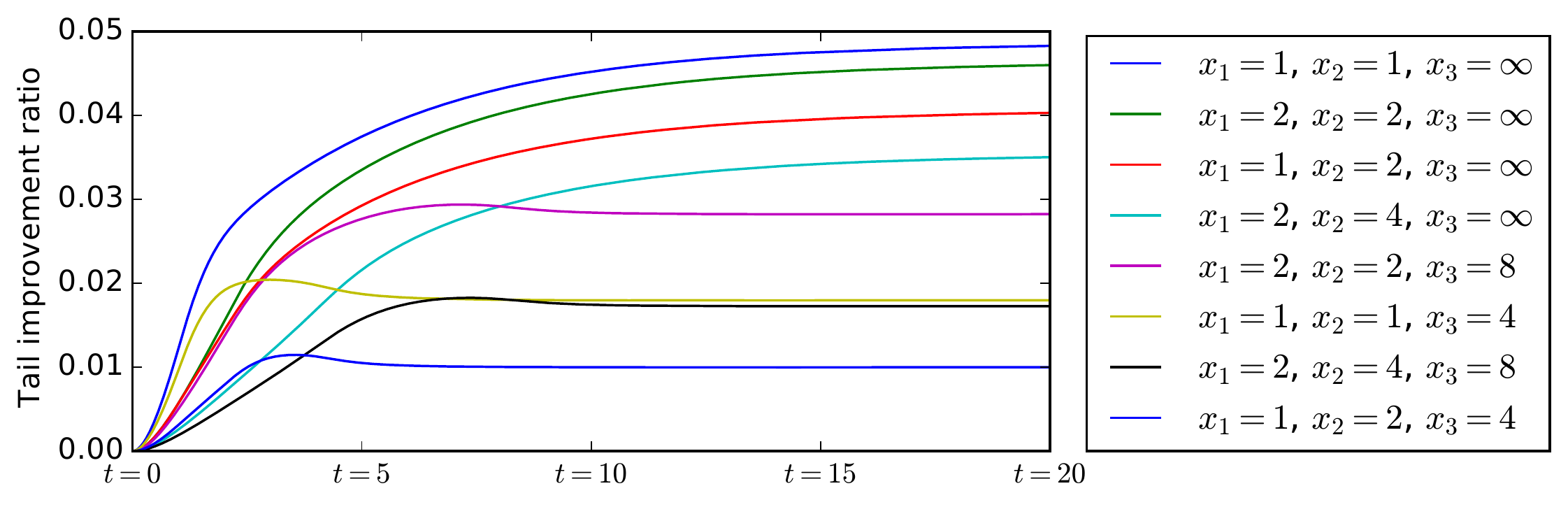}
	    \caption{Empirical tail improvement of Nudge over FCFS under a variety of Nudge parameter choices.
	    Highest improvement occurs when $x_1 = x_2$, $x_3 = \infty$.
	    Job size distribution is hyperexponential with branches drawn from $\Exp(2)$ and $\Exp(1/3)$,
	    where the first branch has probability $0.8$. $\E{S} = 1$, $C^2 = 3$.
	    Simulations run with 10 billion arrivals. Load $\rho = 0.8$.
	    Parameter choices are listed in order of asymptotic improvement.}
	    \label{fig:how_set_nudge_parameters}
	\end{figure}
	
	In \cref{fig:how_set_nudge_parameters}, we show one instance of this pattern, for the case of a particular hyperexponential distribution.   
	We see that the two choices of Nudge parameters
	that display the {\em least} improvement over FCFS
	are those where both medium and very large jobs exist,
	i.e. $x_1 \neq x_2$ and $x_3 \neq \infty$.
	
	To explain this phenomenon, note that when we remove medium and very large jobs, we end up with more swaps.  Empirically, we have found that the \emph{quantity} of swaps
	is more important than the \emph{quality} of those swaps, and thus maximizing the 
	number of swaps leads to the largest improvement.
	While empirically removing medium and very large jobs improves performance,
	our analytical result in \cref{thm:existence_of_improvement}
	requires medium and very large jobs.
	
	Setting the Nudge parameters so that all jobs are either large or small
	dramatically simplifies the problem of choosing Nudge parameters,
	in addition to achieving consistently strong performance.
	Now, only {\em one free parameter} remains: the cutoff between small and large jobs.

	\subsection{Low Load: Dramatic Improvement when Variability is High}
	\label{sec:empirical_low_load}
	
	At low load, Nudge has the potential for dramatic improvement over FCFS (>10\% throughout the tail),
	in the common case where the job size distribution
	is more variable than an exponential distribution,
	i.e. $C^2 > 1$.
	On the other hand,
	under low-variability job size distributions  ($C^2 < 1$),
	we find that Nudge's improvement shrinks at lower loads; here it helps to set the $x_1$ cutoff close to $0$.

 	\begin{figure}[t]
	    \centering
	    \includegraphics[width=\linewidth]{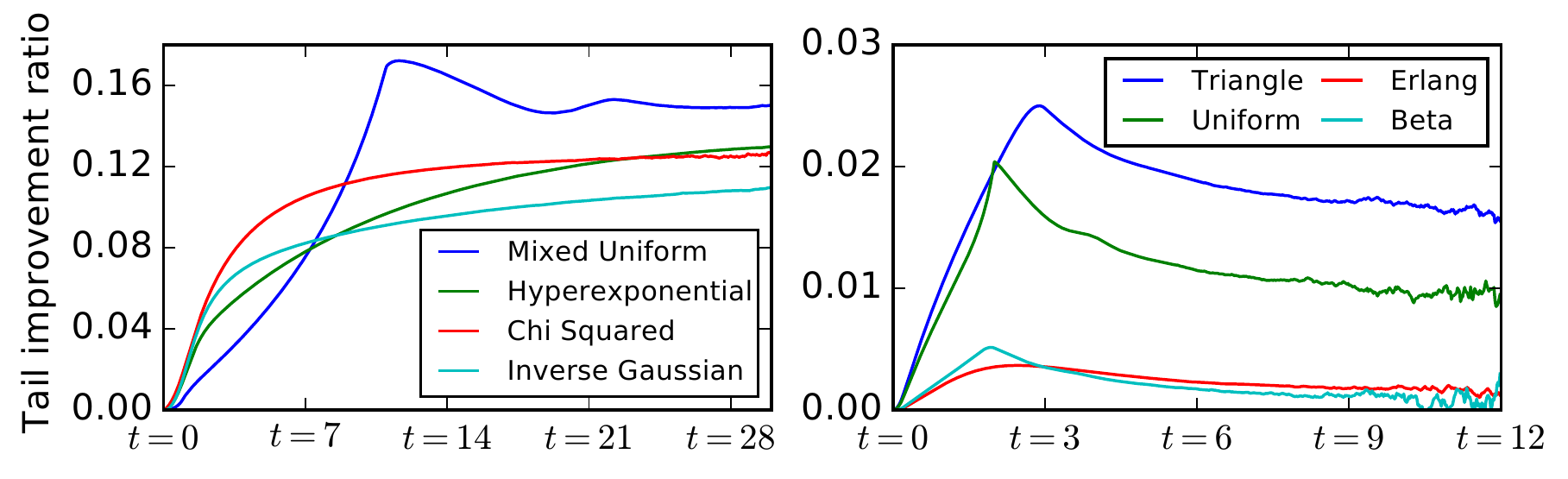}
 	{\small (a) High variance: $C^2 > 1$; $x_1 = x_2 = 1$, $x_3 = \infty$ \hspace{10pt} (b) Low variance: $C^2 < 1$; $x_1 = x_2 = 0.2$, $x_3 = \infty$\vspace{-5pt}}
	    \caption{
	    Empirical tail improvement of Nudge over FCFS at low load ($\rho = 0.4$) under a variety of job size distributions with $\E{S} = 1$.
	    (a) Higher variance distributions show dramatic improvement. (b) Lower variance distributions show modest improvement. Specific distributions: In (a),
	    Mixed Uniform: Uniform($0,1$) w/prob. $0.9$, else Uniform($0,11$), $C^2 = 3.33$;
	    Hyperexponential: $\Exp(2)$ w/ prob. $0.8$, else $\Exp(1/3)$, $C^2 = 3$;
	    ChiSquared(1), $C^2 = 2$;
	    InverseGaussian($\mu=1, \lambda=1/2$), $C^2 = 2$.
	    In (b), Triangle w/ $\min=0$, mode~$=0$, $\max=3$, $C^2 = 1/2$;
	    Uniform($0, 2$), $C^2 = 1/3$;
	    Erlang($k=3$, $\lambda=1/3$, $C^2 = 1/3$;
	    Beta($\alpha = 2, \beta = 2$) scaled by a factor of 2, $C^2 = 1/5$.
	    Distributions listed in order of asymptotic improvement.
	    Simulations run with 10 billion arrivals.}
	    \label{fig:empirical_low_load}
	\end{figure}   

	In \cref{fig:empirical_low_load}
	we show these patterns for a wide variety of distributions
	at relatively low load $\rho = 0.4$.
	In \cref{fig:empirical_low_load}(a), we have four high-variance job size distributions,
	each with $C^2 \in [2, 3.33]$.
	In every case, Nudge dramatically improves upon FCFS, with TIR in the range of 10-15\%.
	In \cref{fig:empirical_low_load}(b), we have four low-variance job size distributions,
	each with $C^2 \in [1/5, 1/2]$.
	In these cases, we reduce the cutoff $x_1$ to $0.2$ for best performance,
	and Nudge's improvement over FCFS is under 3\%.
	
	Intuitively, when load is low, each job waits behind fewer other jobs on average,
	so Nudge's one swap per job has a greater relative impact.
	When those swaps are broadly beneficial for the tail,
	as occurs when job size variance is high,
	Nudge achieves the most dramatic improvement over FCFS.
	When job size variance is low,
	swaps involving small jobs that are near the mean job size
	cause the response time of the large jobs to suffer too much. To alleviate this,
	we reduce the small job cutoff $x_1$
	to maintain stochastic improvement over FCFS.
	
	\subsection{Asymptotic Improvement Means Stochastic Improvement}
	\label{sec:stochastic_asymptotic_connection}
	
	After extensively simulating Nudge under different loads and job size distributions,
	we have found that the space of parameters under which Nudge asymptotically improves upon FCFS
	typically matches the space under which Nudge stochastically improves upon FCFS.

	\begin{table}[t]
	    \centering
	    \setlength\tabcolsep{0.9pt}
	    \begin{tabular}{|c|ccc|ccc|ccc|ccc|ccc|}
	        \hline
	        Job size dist. & $x_1$ & Asym. & Stoc. & $x_1$ & Asym. & Stoc. & $x_1$ & Asym. & Stoc.
	        & $x_1$ & Asym. & Stoc. & $x_1$ & Asym. & Stoc.\\ \hline
	        Exponential    & $0.5$ & \checkmark & \checkmark
	                       & $1$   & \checkmark & \checkmark
	                       & $2$   & $\times$   & $\times$
	                       & $4$   & $\times$   & $\times$
	                       & $8$   & $\times$   & $\times$
	                       \\ \hline
	        Hyperexponential
	                       & $0.5$ & \checkmark & \checkmark
	                       & $1$   & \checkmark & \checkmark
	                       & $2$   & \checkmark & \checkmark
	                       & $4$   & \checkmark & \checkmark
	                       & $8$   & \checkmark & \checkmark
	                       \\ \hline
	        Bounded Lomax
	                       & $0.5$ & \checkmark & \checkmark
	                       & $1$   & \checkmark & \checkmark
	                       & $1.5$ & $\times$   & $\times$
	                       & $2$   & $\times$   & $\times$
	                       & $3$   & $\times$   & $\times$
	                       \\ \hline
	        Uniform
	                       & $0.1$  & \checkmark & \checkmark
	                       & $0.2$  & \checkmark & \checkmark
	                       & $0.5$  & $\times$   & $\times$
	                       & $0.75$ & $\times$   & $\times$
	                       & $1$    & $\times$   & $\times$
	                       \\ \hline
	        Beta
	                       & $0.1$  & \checkmark & \checkmark
	                       & $0.2$  & \checkmark & \checkmark
	                       & $0.3$  & $\times$   & $\times$
	                       & $0.4$  & $\times$   & $\times$
	                       & $0.5$  & $\times$   & $\times$
	                       \\ \hline
	    \end{tabular}
	    \caption{Presence or absence of asymptotic and stochastic improvement of Nudge over FCFS
	    under a variety of job size distributions and Nudge parameter choices.
	    Stochastic improvement occurs whenever asymptotic improvement occurs.
	    Each row gives a distinct job size distribution,
	    and each cell gives a distinct Nudge parameter setting.
	    In every case, $x_2 = x_1$ and $x_3 = \infty$,
	    so only $x_1$ is specified.
	    Load $\rho = 0.4$.
	    Specific job size distributions, each with mean 1: Exponential with mean 1, Uniform(0, 2),
	    Hyperexponential drawn from $\Exp(2)$ w/ prob. $0.8$ and $\Exp(1/3)$ with prob. $0.2$,
	    BoundedLomax($\lambda = 2, \max=4, \alpha=2$),
	    Beta($\alpha=2, \beta=2$) scaled by a factor of 2.}
	    \label{tab:asymptotic_implies_stochastic}
	\end{table}

	In \cref{tab:asymptotic_implies_stochastic},
	we show the consistency of this relationship across a wide variety of
	job size distributions and choices of Nudge parameters.
	The distributions range from a low-variance Beta distribution with $C^2 = 1/5$
	to a hyperexponential distribution with $C^2 = 3$.
	Across the spectrum, Nudge stochastically improves over FCFS whenever it asymptotically improves over FCFS.

	This connection between asymptotic and stochastic improvement
	is surprising given that the conditions that we need to prove stochastic improvement
	(\cref{thm:improvement_regime})
	are much more stringent than what we need to prove asymptotic improvement
	(\cref{thm:asymptotic_constant}).
	Nonetheless, the connection is highly useful because we have provided a simple analytical formula for determining when Nudge asymptotically improves upon FCFS (\cref{thm:asymptotic_constant}).  

    \section{Nudge in practice}
    \label{sec:nudge_in_practice}
    
    Nudge can be used in practice even if some of the assumptions made in this paper are not perfectly satisfied.
    
    In this paper, we assume that exact job size information is known to the scheduler.
    However, such information is only used to determine which size class (small, large, etc.)
    a job should be placed in.
    In practice, only estimates of job size may be known.
    In such a setting, the scheduler could assign jobs
    that are clearly above or below a size threshold to the large and small classes,
    while placing ambiguous jobs in the medium class.
    If the estimates are reasonably accurate,
    we would expect such a Nudge policy to stochastically improve upon FCFS.
    
    We also assume that the exact job size distribution is known to the scheduler.
    This assumption is needed to choose the Nudge parameters for our proofs in \cref{sec:main_results}.
    However, our empirical results in \cref{sec:empirical} show that
    much less information is needed in practice to choose good Nudge parameters.
    For instance, as we saw in \cref{sec:empirical_low_load},
    the following choice of parameters works well empirically:
    By default, set $x_1 = x_2 = E[S], x_3 = \infty$.
    However, if load is low and job size variability ($C^2$) is low,
    set $x_1 = x_2 = E[S]/5, x_3 = \infty$.
    
    \section{Variants on Nudge}
    \label{sec:nudge_variants}
    As Nudge is such a simple policy,
    there are many interesting variants of Nudge that one could investigate.
    We now discuss the advisability of several such variants.
    
    Recall that Nudge only ever swaps a job at most once.
    One might consider allowing a job to swap a second or third time with new arrivals,
    or even an unlimited number of times.
    Unfortunately,
    this change could ruin Nudge's stochastic improvement over FCFS, if implemented poorly.
    In particular, under a Nudge variant where large jobs can be swapped with an unlimited number of small arrivals,
    such highly-swapped large jobs will typically dominate the response time tail,
    dramatically worsening the variant's tail performance.
    A wiser variant might be to allow large jobs to be swapped with a bounded number of small jobs,
    or to allow only the small jobs to be swapped an unlimited number of times.
    
    Another interesting variant of Nudge would only swap in a probabilistic fashion,
    such as with an i.i.d. coin flip.
    We believe such a policy could achieve stochastic improvement over FCFS.
    However, proving such a result would be no simpler than for Nudge,
    because probabilistic swapping does not change the shape of the distribution of swaps.
    Moreover, the variant's tail improvement ratios would likely be smaller than those of Nudge,
    because a smaller fraction of jobs are involved in swaps.
    
    Finally, one could design a more complicated variant of Nudge
    which would consider a job's exact size when deciding whether to swap,
    rather than simply comparing the job's size to a threshold.
    For instance, one might decide to swap all pairs of jobs whose sizes differ by a factor of 2,
    as long as neither job has yet been swapped.
    These more complicated Nudge variants
    might achieve even larger stochastic improvements over FCFS than Nudge.
    Beyond FCFS, such Nudge variants
    might be able to stochastically improve upon some or even all Nudge policies.
    We leave this possibility as an open question.

	\section{Conclusion}
	We introduce Nudge,
	the first scheduling policy
	whose response time distribution stochastically improves upon that of FCFS.
	Specifically, we prove that with appropriately chosen parameters,
	Nudge stochastically improves upon FCFS for light-tailed job size distributions%
	\footnote{%
	More specifically, continuous class I job size distributions with positive density at 0.%
	}.
	From an asymptotic viewpoint,
	we prove that Nudge achieves a multiplicative improvement over FCFS,
	disproving the strong asymptotic optimality conjecture for FCFS.
	Finally, we derive the Laplace-Stieltjes transform of response time under Nudge,
	using a novel technique.
	Nudge is simple to implement
	and is a practical drop-in replacement
	for FCFS when job sizes are known.
	
	One of the major insights of this paper is that 
	improving the tail does not follow
	the same intuitions that we use in improving the mean.
	While improving mean response time is often a matter of
	helping small jobs jump ahead of large ones,
	when it comes to the tail,
	this has to be done in a very measured way.
	Too much help to the small jobs causes the tail to get a lot worse.
	Nudge finds the exactly appropriate way to do this.  
	
	One direction for future work is further exploring the stochastic improvement frontier.
	Can we stochastically improve upon other commonly used scheduling policies?
	Can we improve upon Nudge itself, such as with a more complicated variant of Nudge (see \cref{sec:nudge_variants})?
	One policy which cannot be stochastically improved upon is SRPT,
	due to its optimal mean response time.
	Can we prove that other policies are unimprovable?
	
	Another direction is simplifying the definition of Nudge.
	Our empirical results in \cref{sec:empirical} indicate that in practice,
	Nudge can always stochastically improve upon FCFS with only two classes of jobs: small and large.
	It would be of practical importance to figure out how to extend the theorems in this paper to hold for this simplified definition of Nudge.
	
	\section*{Acknowledgements}
	We thank Sem Borst and the anonymous referees for their helpful comments.
    This research was supported by NSF-CMMI-1938909, NSF-CSR-1763701, and a Google 2020 Faculty Research Award.
	
	\bibliographystyle{ACM-Reference-Format}
	\bibliography{references}
	
	\appendix
	\section{Proofs for Stochastic Improvement}
	\label{app:normalized_converges}
	
	\begin{replemma}{lem:normalized_converges}
	    Suppose $S$ is a continuous class I job size distribution.
	    For any load~$\rho$,
	    the normalized p.d.f. $g(t)$
	    is bounded above and below by positive constants,
	    and $\lim_{t \to \infty} g(t)$ exists.
	\end{replemma}
	
	\begin{proof}
		First we show (following prior work \citep{abate1994waiting,sakurai2004approximating,abate1997asymptotics}) that $\widetilde{T_Q^\fcfs}$ has a simple pole $-\theta^*$ as its rightmost singularity.
		
		We let $-\theta^*$ be the root of the denominator of $\widetilde{T_Q^\fcfs}(s)$, which is
		\begin{equation*}
			\lambda\widetilde{S}(s)-\lambda+s=0
			\iff
			\widetilde{S}(s)=\frac{\lambda-s}{\lambda}.
		\end{equation*}
		Since the left hand $\widetilde{S}(s)$ is convex in $s$\footnote{We have $\widetilde{S}''(s)=\int_{t=0}^{\infty}t^2e^{-st}f_S(t)\d t>0$ for every $s$ in the convergence region of $\widetilde{S}(\cdot)$.}, and the right hand $\frac{\lambda-s}{\lambda}$ is only linear in $s$, their intersection $s=-\theta^*$ must be a simple root. Moreover, such an intersection exists for $s<0$ because
		\begin{itemize}
			\item $\widetilde{S}(0)=\frac{\lambda-0}{\lambda}=1$;
			\item 
			$\widetilde{S}'(0)=-\frac{1}{\mu}>-\frac{1}{\lambda}$;
			\item $\widetilde{S}(s)\to\infty$ when $s$ approaches the rightmost singularity of $\widetilde{S}$ (since $S$ is a class I distribution).
		\end{itemize}  

		Now we use final value theorem to establish the limit of the ratio between the p.d.f.~$f_{T_Q^{\fcfs}}$ and the exponential function $e^{-\theta^*t}$. Recall the function $g(t)=f_{T_Q^\fcfs}(e)e^{\theta^*t}$ and consider its Laplace transform $\widetilde{G}(s)=\widetilde{T_Q^\fcfs}(s-\theta^*)$. Since the poles of $\widetilde{G}(s)$ map one-to-one to the poles of $\widetilde{T_Q^\fcfs}(s-\theta^*)$,
		the above arguments show that every pole of $\widetilde{G}(s)$ is either in the open left half plane or at the origin, and the origin is a simple pole. Therefore, the Final Value Theorem for $g(t)$ tells us
		
		\begin{align}
			\lim_{t\to\infty} f_{T_Q}^{\fcfs}(t) e^{\theta^* t}
			=\lim_{t\to\infty}g(t)
			=\lim_{s\to0}s\widetilde{G}(s)
			=\frac{(1-\rho)\theta^*}
			{-\lambda \widetilde{S}'(-\theta^*)-1} \triangleq g^*>0.
			\label{eq: exponential pdf}
		\end{align}			
		
		Since the limit $g^*$ exists, for $\epsilon=\frac{g^*}{2} $, there exists $N_{\epsilon}<\infty$ such that $\forall t\ge N_\epsilon$, 
		\begin{equation*}
			\left|g(t)-g^*\right|\le \frac{g^*}{2} 
			\ \Rightarrow\ 
			\frac{g^*}2 e^{-\theta^* t}\le f_{T_Q}^{\fcfs}(t)\le \frac{3g^*}2 e^{-\theta^* t}.
		\end{equation*}
		
		Next, we want to show that
		\begin{align}
			\min_{0\le t<N_\epsilon}g(t) > 0 \quad
			\text{ and } \quad
			\max_{0\le t<N_\epsilon}g(t) < \infty.
		\end{align}
		
		First, note that $f_{T_Q}^{\fcfs}$ satisfies the following level-crossing differential equations \citep{brill2000brief} (we abbreviate $f_{T_Q}^{\fcfs}$ to $f$):
		\begin{equation*}
			\begin{cases}
				f(0)=(1-\rho)\lambda;\\
				{f'}(t)=\lambda f(t)-(1-\rho)\lambda f_S(t)-\lambda\int_{j=0}^{t}f(t-j)f_S(j)\d j.
			\end{cases}
		\end{equation*}
		
		To begin with, $f(t)$ is continuous because $f_S(t)$ is continuous.
		If $f(t)=0$ for some $t<N_\epsilon$, we let $t_0=\inf\left\{t<N_\epsilon:f(t)=0\right\}$.
		Clearly $t_0>0$ because $f(0)=(1-\rho)\lambda>0$.
		Note also that $f(t_0) = 0$, because $f$ is continuous.
		Since $f(t_0)<f(0)$, $\exists 0<s_0<t_0$ s.t. $f_S(s_0)>0$. 
		Then $\exists a,b$ where $0\le a<s_0<b\le t_0$, s.t. $f(t)>0$ for all $t\in[a,b]$.
		Now we have 
		\begin{equation*}
			{f'}(t_0)=-(1-\rho)\lambda f_S(t_0)-\lambda\int_{j=0}^{t_0}f(t_0-j)f_S(j)\d j<0
		\end{equation*}
		because the first term $-(1-\rho)\lambda f_S(t_0)\le0$ and the second term
		\begin{equation*}
			-\lambda\int_{j=0}^{t_0}f(t_0-j)f_S(j)\d j
			\le -\lambda\int_{j=a}^b f(t_0-j)f_S(j)\d j
			<0.
		\end{equation*}
		But ${f'}(t_0)<0$ is impossible, because we assumed that $f(t_0) = 0$.
		The implication that ${f'}(t_0)<0$
		contradicts the fact that $f$ is a non-negative probability density function. Therefore, 
		\begin{equation*}
			\min_{0\le t<N_\epsilon} g(t)
			\ge \min_{0\le t<N_\epsilon} g(t)
			>0.
		\end{equation*}
		
		On the other hand, since $f'(t)\le \lambda f(t)$ everywhere, we have $f(t)\le (1-\rho)\lambda e^{\lambda t}$. Note that this bound holds even if $S$ has infinite density at one or more points.
		As a result,
		\begin{equation*}
			\max_{0\le t<N_\epsilon}g(t)\le
			\max_{0\le t<N_\epsilon}\left[f_{T_Q}^{\fcfs}(t)
			e^{\theta^* N_\epsilon}\right]
			\le (1-\rho)\lambda e^{\lambda N_\epsilon}
			e^{\theta^* N_\epsilon}<\infty.
		\end{equation*}

		Finally, note that 
		\begin{align*}
		\inf_{t\in[0, \infty)} g(t) &\triangleq g_{\min}\ge\min\left\{\inf_{0\le t<N_\epsilon}\left[f_{T_Q}^{\fcfs}(t)e^{\theta^* t}\right], \frac{g^*}{2}\right\}>0\quad \\
			\sup_{t\in[0, \infty)} g(t) &\triangleq g_{\max}\le\max\left\{\sup_{0\le t<N_\epsilon}\left[f_{T_Q}^{\fcfs}(t)e^{\theta^* t}\right], \frac{3g^*}{2}\right\}<\infty
		\end{align*}
		which indicates both $g_{\min}$ and $g_{\max}$ are well-defined and nonzero.
		This completes the proof.
	\end{proof}
	
	\begin{replemma}{lem:A_B_c}
		Let $A,B$ be two independent real-valued random variables and $c$ be a fixed constant.
		Suppose $1\le A\le c$ and $A<B$. Under these assumptions, if $\P{B>c}>0$ and
		\begin{align}
			c\E{B}\ge \E{A}\E{B|B>c},
			\label{ineq: A B c condition}
		\end{align}
		then
		\begin{align}
			\frac{\E{\min(AB, c)- \min(A, c)}}{\E{\min(AB, c)-\min(B, c)}}
			= \frac{\E{\min(AB, c)-A}}{\E{\min(AB, c)-\min(B, c)}}
			\ge\frac{\E{AB-A}}{\E{AB-B}}.
			\label{ineq: A B c}
		\end{align}
	\end{replemma}
	
	\begin{proof}
		First we observe
		${\E{\min(AB, c)-A}}>{\E{\min(AB, c)-\min(B, c)}}$ because $A<\min(B, c)$.
		Based on this, we can shrink the left hand side of inequality~\eqref{ineq: A B c} by adding the same positive term to both the denominator and numerator.
		We compute
		\begin{align}
		    \label{eq:tmp1}
			\frac{\E{\min(AB, c)-A}}{\E{\min(AB, c)-\min(B, c)}}&=
			\frac{\E{AB-A}-\E{(AB-c)\indict{AB>c}}}
			{\E{AB-B}-\E{(AB-c)\indict{AB>c}}+\E{(B-c)\indict{B>c}}}.
		\end{align}
		Since $A\ge1$ and $B>0$, we have $AB\ge B$. Therefore,
		\begin{equation*}
			0\le \indict{AB>c}-\indict{B>c}\le \indict{AB>c}=\indict{AB-c>0}.
		\end{equation*} 
		We proceed by adding a positive term, $\E{(AB-c)\left(\indict{AB>c}-\indict{B>c}\right)}$,
		to both the denominator and numerator of the right hand side of \eqref{eq:tmp1} and obtain
		\begin{align*}
			\frac{\E{\min(AB, c)-A}}{\E{\min(AB, c)-\min(B, c)}}
			&\ge
			\frac{\E{AB-A}-\E{(AB-c)\indict{B>c}}}
			{\E{AB-B}-\E{(AB-B)\indict{B>c}}}.
		\end{align*}
		Hence, to establish inequality~\eqref{ineq: A B c}, it suffices to show
		\begin{align*}
			&\frac{\E{(AB-c)\indict{B>c}}}
			{\E{(AB-B)\indict{B>c}}}\ge
			\frac{\E{AB-A}}{\E{AB-B}}\\
			\iff&\E{(AB-c)\indict{B>c}}\E{AB-B}\ge
			\E{(AB-B)\indict{B>c}}\E{AB-A}\\
			\iff&
			\left(\E{A}\E{B|B>c}-c\right)\P{B>c}\E{AB-B}
			\ge\left(\E{A}-1\right)
			\E{B|B>c}
			\P{B>c}\E{AB-A}\\
			\iff&
			c\E{B}\left(\E{A}-1\right)\ge 
			\E{A}\E{B|B>c}\left(\E{A}-1\right)\\
			\iff&
			c\E{B}\ge 
			\E{A}\E{B|B>c},
		\end{align*}
		which is precisely the condition provided in \eqref{ineq: A B c condition}.	
	\end{proof}

	\section{Proofs for Transform Analysis}
	\label{app:transform_proofs}
	
    \begin{replemma}{lem:swap_fcfs}
        The small tagged job swaps with a large job in the Nudge system
        if and only if, when it arrives,
        the FCFS system has a nonempty queue whose last job is large.
    \end{replemma}
    
    \begin{proof}
        By definition of Nudge, the tagged job swaps if and only if, when it arrives,
        the Nudge system has a nonempty queue whose last job is
        a large job that has not been swapped.
        It therefore suffices to show that at any moment in time,
        the FCFS system has a nonempty queue whose last job is large
        if and only if the Nudge system has a nonempty queue
        whose last job is a large job that has not been swapped.
        
        We first note that the total amount of work in both systems
        is the same at every moment in time,
        because both FCFS and Nudge are work conserving.
        
        Suppose the FCFS system has a nonempty queue
        whose last job~$j$ is large.
        Because it is the last job in the FCFS queue,
        there have been no new arrivals after~$j$.
        In the Nudge system, this means $j$ has not been swapped,
        so either $j$ is the last job in the Nudge queue or has entered service.
        By work conservation, both systems had the same amount of work when $j$ arrived,
        so $j$ must still be in the Nudge queue, as desired.
        
        Suppose the Nudge system has a nonempty queue
        whose last job~$j$ is a large job that has not been swapped.
        We argue similarly to the previous direction:
        there have been no arrivals since~$j$
        because it is at the end of the Nudge queue without being swapped,
        and $j$ cannot have entered service in the FCFS system
        by work conservation,
        so $j$ must be the last job in the FCFS queue, as desired.
    \end{proof}
	
	\begin{replemma}{lem:lst_small_q}
        Let $q \triangleq \P{\nop{W}{s} \ \land\ N_Q = 0}$.
        We have
        \begin{equation*}
            q
            = \widetilde{T_Q^\fcfs}(\lambda) \cdot \frac{\lambda \widetilde{S}(s) - s \widetilde{S}(\lambda)}{\lambda - s} \\
            = \frac{1 - \rho}{\widetilde{S}(\lambda)} \cdot \frac{\lambda \widetilde{S}(s) - s \widetilde{S}(\lambda)}{\lambda - s}.
        \end{equation*}
	\end{replemma}
	
	To prove \cref{lem:lst_small_q}, we require an additional lemma.
	
    \begin{lemma}
        \label{lem:lst_two_exp}
        Let $V$ be a nonnegative random variable,
        and let $\Exp(r)$ and $\Exp(s)$ be exponentially distributed random variables
        of rates $r$ and~$s$, respectively.
        Suppose $V$, $\Exp(r)$, and $\Exp(s)$ are mutually independent.
        Then
        \begin{equation*}
            \P{V < \Exp(r) + \Exp(s)} = \frac{r \widetilde{V}(s) - s \widetilde{V}(r)}{r - s}.
        \end{equation*}
    \end{lemma}
    
    \begin{proof}
        We compute
        \begin{align*}
            \P{V < \Exp(r) + \Exp(s)}
            &= \int_{v = 0}^\infty \P{v < \Exp(r) + \Exp(s)} f_V(v) \d{v} \\
            &= \int_{v = 0}^\infty \left( \int_{u = 0}^\infty \int_{t = 0}^\infty
                \mathbf{1}(v < t + u) \cdot r e^{-rt} \cdot s e^{-su} \d{t}\d{u} \right) f_V(v) \d{v} \\
            &= \int_{v = 0}^\infty \frac{r e^{-rv} - s e^{-sv}}{r - s} f_V(v) \d{v} \\
            &= \frac{r \widetilde{V}(s) - s \widetilde{V}(r)}{r - s}.
            \qedhere
        \end{align*}
    \end{proof}
	
    \begin{proof}[Proof of \cref{lem:lst_small_q}]
        Consider a FCFS system in equilibrium
        along with an independent Poisson ``interruption'' process of rate~$s$.
        Call a job \emph{lucky} if it enters the system while $N_Q = 0$
        and experiences no interruptions during its queueing time.
        Because Poisson arrivals see time averages \citep{wolff_poisson_1982},
        $q$ is probability an arriving job is lucky.
        
        We compute $q$ in an unusual way.
        Let a job's \emph{departure period}
        be the time interval starting when the job enters service
        and ending when the next job enters service.
        Jobs enter service at average rate~$\lambda$,
        so $q$ is the average number of lucky jobs that arrive a departure period.
        More formally, by renewal-reward theorem,
        \begin{align*}
            q
            &= \P{\text{arrival is lucky}} \\
            &= \frac{\text{average rate of lucky arrivals}}{\lambda} \\
            &= \frac{\text{average rate of lucky arrivals}}{\text{average rate of departure periods}} \\
            &= \E{\text{number of lucky arrivals during a departure period}}.
        \end{align*}
        Moreover, because a job is lucky only if $N_Q = 0$,
        only the first arrival in a departure period can possibly be lucky, so
        \begin{equation*}
            q = \P{\text{first arrival in a departure period is lucky}}.
        \end{equation*}
        
        Consider a job~$j$.
        The first arrival in $j$'s departure period is lucky
        if both of the following events occur:
        \begin{align*}
            E_1 &= \text{there are no arrivals during $j$'s queueing time} \\
            E_2 &= \text{$j$ completes before the first interruption after the first arrival of $j$'s departure period}.
        \end{align*}
        By \cref{eq:lst_nop}, $\P{E_1} = \widetilde{T_Q^\fcfs}(\lambda)$.
        We compute $\P{E_2 \mid E_1}$ below.
        
        Conditioned on~$E_1$, the queue is empty when $j$ enters service,
        so the first arrival during $j$'s departure period is simply
        the first arrival after $j$ enters service.
        Let $\Exp(\lambda)$ be the amount of time between when $j$ enters service
        and the next arrival,
        and let $\Exp(s)$ be the amount of time between that next arrival
        and the first interruption after it.
        Both $\Exp(\lambda)$ and $\Exp(s)$ are exponentially distributed
        with rates $\lambda$ and~$s$, respectively,
        and they and $j$'s size are mutually independent.
        Because $j$'s size is distributed as~$S$, we have
        \begin{equation*}
            \P{E_2 \mid E_1}
            = \P{S < \Exp(\lambda) + \Exp(s)}
            = \frac{\lambda \widetilde{S}(s) - s \widetilde{S}(\lambda)}{\lambda - s},
        \end{equation*}
        where the latter equality follows from \cref{lem:lst_two_exp} below.
        
        It remains only to show
               $\widetilde{T_Q^\fcfs}(\lambda) = (1 - \rho)/\widetilde{S}(\lambda)$.
        Because response time $T^\fcfs$ is a sum of independent random variables
        with distributions $T_Q^\fcfs$ and~$S$, we have
        $\widetilde{T_Q^\fcfs}(\lambda) = \widetilde{T^\fcfs}(\lambda)/\widetilde{S}(\lambda)$.
        By \eqref{eq:lst_nop},
        $\widetilde{T^\fcfs}(\lambda)$ is the probability that
        no arrivals occur during a job's response time.
        But this is simply the probability that a job leaves an empty system when it departs,
        which is $1 - \rho$.
    \end{proof}
    
	\begin{reptheorem}{thm:asymptotic_constant}
		Suppose $S$ is a continuous class I job size distribution.
		For any $s_{\min} < x_1 \leq x_2 \leq x_3$,
		the asymptotic tail improvement ratio
		of $\nudge(x_1, x_2, x_3)$ compared to $\fcfs$ is
		\begin{equation*}
			\asymtir =
			\psmall\plarge \frac{\lambda}{\lambda+\theta^*}\left(
				\widetilde{\slarge}(-\theta^*)
				-\frac{\lambda}{\lambda+\theta^*}
				\widetilde{\ssmall}(-\theta^*)
				-\frac{\theta^*}{\lambda+\theta^*}
				\widetilde{\slarge}(-\theta^*)\widetilde{\ssmall}(-\theta^*)
			\right).
		\end{equation*}
		Furthermore, $\asymtir$ is positive, meaning $C_\nudge < C_\fcfs$, if
		\begin{equation*}
			\frac{\lambda+\theta^*}{\lambda}<
			\frac{1-\widetilde{\slarge}(-\theta^*)^{-1}}
			{1-\widetilde{\ssmall}(-\theta^*)^{-1}}.
		\end{equation*}
	\end{reptheorem}
    
    \begin{proof}
	We prove this theorem using the Laplace-Stieltjes transform derived in \cref{thm:transform}.
	
	The transform of the tail of $T^{\nudge}$ can be calculated as
	\begin{align*}
		\int_{t=0}^{\infty}
		e^{-st}\P{T^\nudge>t}\d t
		&=-\int_{t=0}^{\infty}
		\P{T^\nudge>t}\d \left(\frac{e^{-st}}{s}\right)\\
		&=-\left.\P{T^\nudge>t}\left(\frac{e^{-st}}{s}\right)\right|_0^{\infty}
		+\int_{t=0}^{\infty}\frac{e^{-st}}{s}\d\P{T^\nudge>t}\\
		&=\frac{1}{s}\left(1-
		\int_{t=0}^{\infty} e^{-st}f_{T^{\nudge}}(t)\d t\right)\\
		&=\frac{1-\widetilde{T^{\nudge}}(s)}{s}.
		\yestag
		\label{eq:nudge_tail_transform}
	\end{align*}

	Then the transform of $e^{\theta^*t}\P{T^\nudge>t}$ is obtained by translating \cref{eq:nudge_tail_transform} horizontally through $\theta^*$ units:
	\begin{equation*}
	    \int_{t=0}^{\infty} e^{-st} \left( e^{\theta^*t}\P{T^\nudge>t} \right) \d t = \frac{1-\widetilde{T^{\nudge}}(s-\theta^*)}{s-\theta^*}.
	\end{equation*}
    Now, we are ready to calculate $C_\nudge$ using this transform.
	From Final Value Theorem,
	\begin{align*}
		C_\nudge&=\lim_{t\to\infty} e^{\theta^*t}\P{T^\nudge>t}\\
		&=\lim_{s\to0}s\frac{1-\widetilde{T^{\nudge}}(s-\theta^*)}{s-\theta^*}\\
		&=\frac{1}{\theta^*}\lim_{s\to0}s\widetilde{T^{\nudge}}(s-\theta^*)
	\end{align*}
	Now, we substitute the expression from \cref{thm:transform},
	and drop terms that are negligible in $s \to 0$ limit.
	\begin{align*}
	    &\frac{1}{\theta^*}\lim_{s\to0}s\widetilde{T^{\nudge}}(s-\theta^*) \\
		&=C_\fcfs+\frac{1}{\theta^*}\psmall\plarge\left(
			\widetilde{\slarge}(-\theta^*)
			(\widetilde{\ssmall}(-\theta^*)-1)C_{Q,\fcfs}
			+\widetilde{\ssmall}(-\theta^*)
			(1-\widetilde{\slarge}(-\theta^*))\frac{C_{Q,\fcfs}}{\widetilde{S}(-\theta^*)}
		\right),
		\yestag
		\label{eq:c_nudge_before_simplifying}
	\end{align*}
	where 
	\begin{equation}
		C_{Q,\fcfs}=\lim_{s\to 0}s\widetilde{T_Q^\fcfs}(s-\theta^*)
		=\lim_{s\to 0}s
		\frac{\widetilde{T^\fcfs}(s-\theta^*)}
		{\widetilde{S}(s-\theta^*)}=\frac{\theta^* C_\fcfs}{\widetilde{S}(-\theta^*)}.
		\label{eq:c_q_fcfs}
	\end{equation}
    
	We recall that $-\theta^*$ is the rightmost singularity of $\widetilde{T_Q^\fcfs}(s)=\frac{(1-\rho)s}{\lambda\widetilde{S}(s)-\lambda+s}$, which indicates that $\theta^*$ is the smallest positive value that satisfies
	\begin{equation}
		\lambda\widetilde{S}(-\theta^*)-\lambda-\theta^*=0
		\quad \text{ and } \quad
		\widetilde{S}(-\theta^*)=\frac{\lambda+\theta^*}{\lambda}.\label{eq:tilde_s_-theta}
	\end{equation}
	Using \cref{eq:c_q_fcfs} and \cref{eq:tilde_s_-theta} to simplify \cref{eq:c_nudge_before_simplifying}, we obtain
	\begin{align*}
		&C_\nudge \\
		&=
		C_\fcfs\left(1-\psmall\plarge \frac{\lambda}{\lambda+\theta^*}\left(
			\widetilde{\slarge}(-\theta^*)
			-\frac{\lambda}{\lambda+\theta^*}
			\widetilde{\ssmall}(-\theta^*)
			-\frac{\theta^*}{\lambda+\theta^*}
			\widetilde{\slarge}(-\theta^*)\widetilde{\ssmall}(-\theta^*)\right)\right).
	\end{align*}
	This gives us
	\begin{align*}
		\asymtir&=1-\frac{C_\nudge}{C_\fcfs}\\
		&=\psmall\plarge \frac{\lambda}{\lambda+\theta^*}\left(
			\widetilde{\slarge}(-\theta^*)
			-\frac{\lambda}{\lambda+\theta^*}
			\widetilde{\ssmall}(-\theta^*)
			-\frac{\theta^*}{\lambda+\theta^*}
			\widetilde{\slarge}(-\theta^*)\widetilde{\ssmall}(-\theta^*)
			\right).
	\end{align*}
	By assumption,
	\begin{equation*}
		\frac{\lambda+\theta^*}{\lambda}<
			\frac{1-\widetilde{\slarge}(-\theta^*)^{-1}}
			{1-\widetilde{\ssmall}(-\theta^*)^{-1}},
	\end{equation*}
	 so we have
	 \begin{align*}
		 \asymtir&=\psmall\plarge \frac{\lambda}{\lambda+\theta^*}
		 \widetilde{\slarge}(-\theta^*)\widetilde{\ssmall}(-\theta^*)
		 \left(
			\widetilde{\ssmall}(-\theta^*)^{-1}
			-\frac{\lambda}{\lambda+\theta^*}
			\widetilde{\slarge}(-\theta^*)^{-1}
			-\frac{\theta^*}{\lambda+\theta^*}
		\right)	\\
		&=\psmall\plarge \frac{\lambda}{\lambda+\theta^*}
		\widetilde{\slarge}(\!-\theta^*\!)\widetilde{\ssmall}(\!-\theta^*\!)
		\left(
			\frac{\lambda}{\lambda\!+\!\theta^*}
			\left(\!1\!-\!\widetilde{\slarge}(-\theta^*)^{-1}\!\right)
			-\left(\!1\!-\!\widetilde{\ssmall}(-\theta^*)^{-1}\!\right)
	   \right)
	   >0.
	 \end{align*}
	 Hence $C_\nudge<C_\fcfs.$
	\end{proof}
\end{document}